\documentclass[aps,pra,twocolumn,showpacs,superscriptaddress]{revtex4-1}

\usepackage{amsmath}  
\usepackage{amsthm}  
\usepackage{amssymb}  
\usepackage{color}
\usepackage{longtable}
\usepackage{multirow}
\usepackage{mathrsfs}
\usepackage{braket}
\usepackage{xfrac}
\usepackage[T1]{fontenc}

\let\savedegree\corresponds
\let\corresponds\relax
\usepackage{mathabx}
\let\corresponds\savedegree

\usepackage{float}
\usepackage{bm}
\usepackage[percent]{overpic}
\usepackage{mathrsfs}
\usepackage{mathcomp}
\usepackage[table,dvipsnames]{xcolor}

\usepackage[caption=false]{subfig}
\usepackage{graphicx}

\begin{document}
\title{Deriving quantum constraints and tight uncertainty relations}

\author{Arun Sehrawat}
\email[Email: ]{arunsehrawat@hri.res.in}
\affiliation{Department of Physical Sciences, 
	Indian Institute of Science Education \& Research Mohali, 
	Sector 81 SAS Nagar, Manauli PO 140306, 
	Punjab, India}
\altaffiliation[Current address: ]
{Harish-Chandra Research Institute, Chhatnag Road, Jhunsi, Allahabad 211019, India.}


\begin{abstract}
We present a systematic procedure to obtain all necessary and sufficient (quantum) constraints on the expectation values for any set of qudit's operators.
These constraints---arise form Hermiticity, normalization, and
positivity of a statistical operator and through Born's rule---analytically define an allowed region.
A point outside the admissible region does not correspond to any quantum state, whereas every point in it come from a quantum state.
For a set of observables, the allowed region is a compact and convex set in a real space, and all its extreme points come from pure quantum states.
By defining appropriate concave functions on the permitted region and
then finding their absolute minimum at the extreme points, we obtain different tight uncertainty relations for qubit's and spin observables.
In addition, quantum constraints are explicitly given for the Weyl operators and the spin
observables.
\end{abstract}

\maketitle

\section{Introduction}\label{sec:Intro}

Von Neumann described a
state for a quantum system with
a density (statistical) operator on the system's Hilbert space
\cite{von-Neumann27,von-Neumann55,Fano57}.
A valid density operator must be \emph{Hermitian}, \emph{positive semi-definite}, and of \emph{unit trace}.
Born provided a rule \cite{Born26,Wheeler83} to compute the expectation values for any set of operators from a given statistical operator.
Naturally, all necessary and sufficient constraints---called \emph{quantum constraints} (QCs)---on the expectation values emerge from the three conditions on a density operator.

In Sec.~\ref{sec:QC}, a systematic procedure to derive the QCs is presented, where a result from \cite{Kimura03,Byrd03} is used for the positivity of a statistical operator (or simply a state).
To transfer the conditions from a state onto
the expectation values, one needs 
the Born rule and
an operator-basis to represent operators. 
One can choose any basis, the procedure in Sec.~\ref{sec:QC} is basis independent.

In~\cite{Kimura03,Byrd03}, generators of the special unitary group---that with the identity operator constitute an orthogonal operator-basis---are utilized, and the QCs on their average values are achieved by applying the Lie algebra.
Alternatively, one can start with an orthonormal basis of
the system's Hilbert space, and with all possible ``ket-bra'' pairs one can 
assemble a \emph{standard} operator-basis.
Then, one can exploit the matrix mechanics---developed by Heisenberg, Born, Jordan, and Dirac \cite{Heisenberg25,Born25,Born25-b,Dirac25,Waerden68}---to reach the QCs as demonstrated in Sec.~\ref{sec:QC}.

The QCs and uncertainty relations (URs) are two main strands of this paper.
Heisenberg pioneered the first UR \cite{Heisenberg27,Wheeler83} 
for the position and momentum operators.
A general version of Heisenberg's relation for a pair of operators is introduced by Robertson \cite{Robertson29} that is then improved by
Schr\"{o}dinger \cite{Schrodinger32}.
Deutsch \cite{Deutsch83}, Kraus \cite{Kraus87}, Maassen and
Uffink \cite{Maassen88} formulated URs by employing entropy---rather than the standard deviation that is exercised in \cite{Robertson29,Schrodinger32}---as a measure of uncertainty.
For an overview, we point to \cite{Wehner10,Bialynicki11,Coles17} for entropy URs and \cite{Folland97,Busch07,Busch14} are more in the spirit of Heisenberg's UR.

Throughout the article, we are considering a $d$-level quantum system (qudit).
For a set of \textsc{n} observables (Hermitian operators), 
the QCs bound an \emph{allowed} region $\mathcal{E}$ of the expectation values 
in the real space $\mathbb{R}^{\textsc{n}}$.
If one defines a suitable concave function on $\mathcal{E}$
to measure a combined uncertainty as described in Sec.~\ref{sec:QC}, then
creating a \emph{tight} UR becomes an optimization problem where at most ${2(d-1)}$ parameters are involved (for example, see \cite{Sehrawat17,Riccardi17}).
A UR is called tight if there exists a quantum state that
saturates it.
With this, we close Sec.~\ref{sec:QC} and try its results in the subsequent sections.

In Sec.~\ref{sec:unitary-basis}, we apply the 
general methodology of Sec.~\ref{sec:QC} to the \emph{unitary} operator basis, which is known due to Weyl and Schwinger~\cite{Weyl32,Schwinger60}.
In the case of a prime (power) dimension $d$, the unitary-basis can be divided into ${d+1}$ disjoint subsets such that all the operators in each subset possess a common eigenbasis \cite{Bandyopadhyay02,Englert01}.
These ${d+1}$ eigenbases form a maximal set of mutually unbiased bases (MUBs) \cite{Ivanovic81,Wootters89,Durt10} of the Hilbert space.
In Sec.~\ref{sec:unitary-basis}, QCs 
for the Weyl operators as well as for MUBs are presented.
There we arrive at the same \emph{quadratic} QC that is conceived in \cite{Larsen90,Ivanovic92,Klappenecker05}.
Using the quadratic QC, tight URs for the MUBs are achieved in  \cite{Sanchez-Ruiz95,Ballester07}, and their 
minimum uncertainty states are reported in \cite{Wootters07,Appleby14}.

In the case of ${d\geq3}$, there also exists a \emph{cubic} QC.
In Sec.~\ref{sec:qutrit}, ${d=3}$, QCs
are explicitly given for the Weyl operators of a qutrit and for a set of spin-1 operators.
In addition, a number of tight URs and certainty relations (CRs) are delivered for the spin operators.
By the way, the QCs for the spin-1 operators can also be achieved from \cite{Kimura03,Byrd03}.
In Sec.~\ref{sec:spin-j}, tight URs and CRs are obtained
for the angular momentum operators ${J_x,J_y,}$ and $J_z$, where the quantum number $\mathsf{j}$
can be
${\tfrac{1}{2}, 1,\tfrac{3}{2},2,\cdots\,}$.
The paper is concluded in Sec.~\ref{sec:conclusion}, where a list of our main contributions is prepared.

Appendix~\ref{sec:qubit}
offers a comprehensive analysis for a qubit ${(d=2)}$ that
includes the Schr\"{o}dinger UR \cite{Schrodinger32}, and URs for the 
symmetric informationally complete positive operator valued measure (SIC-POVM) \cite{Rehacek04,Appleby09} are presented there.
In the case of a qubit, it is a known result that $\mathcal{E}$ will be an ellipsoidal region for any number of observables (measurement settings) \cite{Kaniewski14}, and it is also manifested here.
Appendixes~\ref{subsec:2settings} and \ref{subsec:3settings} separately deal with two and three measurement settings.
In the case of two settings, the ellipsoid transfigures into an ellipse,
which also appears in \cite{Lenard72,Larsen90,Kaniewski14,Abbott16,Sehrawat17}.
It is revealed in \cite{Sehrawat17} that several tight CRs and URs known from \cite{Larsen90,Busch14-b,Garrett90,Sanchez-Ruiz98,Ghirardi03,Bosyk12,Vicente05,Zozor13,Deutsch83,Maassen88,Rastegin12}
can be achieved by exploiting the ellipse.
In this article, we deal with Hilbert space $\mathscr{H}_d$ of kets
and Hilbert-Schmidt space $\mathscr{B}(\mathscr{H}_d)$ of operators, 
and their bases are differently symbolized by $\mathcal{B}$ and $\mathfrak{B}$, respectively, to avoid any confusion.

\section{Quantum constraints, allowed region, and uncertainty measures}\label{sec:QC}

Quantum state for a qudit can be described by a statistical operator $\rho$ \cite{von-Neumann27,von-Neumann55,Fano57,Bengtsson06}, on the system's Hilbert space $\mathscr{H}_d$, such that
\begin{eqnarray}
\label{Herm-rho}
\rho&=&\rho^\dagger\quad\ \; \text{(Hermiticity)}\,,\\
\label{norm-rho}
\text{tr}(\rho)&=&1\qquad \text{(normalization)}\,,\qquad\text{and}\\
\label{pos-rho}
0&\leq& \rho\qquad\text{(positivity)}\,.
\end{eqnarray}
The dagger $\dagger$ denotes the adjoint.
It has been shown in \cite{Kimura03,Byrd03} that
an operator $\rho$ fulfills \eqref{pos-rho} if and only if it obeys 
\begin{eqnarray}
\label{0<=S_n}
0&\leq&S_n \quad \mbox{for all} \quad 
1\leq n\leq d\,,
\qquad\quad\text{where}\\
\label{S_n}
S_n&=&\tfrac{1}{n}
\sum_{m=1}^{n}(-1)^{m-1}\,
\text{tr}(\rho^{m})\,S_{n-m}
\end{eqnarray}
commencing with
${S_1=\text{tr}(\rho),}$ and ${S_0:=1}$.
It is advantageous to use inequalities \eqref{0<=S_n} between real numbers
than a single operator-inequality \eqref{pos-rho}; see also \cite{+veMat}.
Due to normalization~\eqref{norm-rho}, the first condition
${0\leq S_1=1}$ holds naturally. 
In a nutshell, an operator $\rho$ on $\mathscr{H}_d$ represents a legitimate quantum state if and only if it complies with 
\eqref{Herm-rho}, \eqref{norm-rho}, and \eqref{0<=S_n} for ${2\leq n\leq d}$.

The set of all bounded operators on $\mathscr{H}_d$ form a $d^2$-dimensional Hilbert-Schmidt space $\mathscr{B}(\mathscr{H}_d)$ endowed with
the inner product
\begin{equation}
\label{HS-inner-pro}
\lgroup A,B\,\rgroup_\textsc{hs}=
\text{tr}(A^\dagger B)\,,
\quad\mbox{where}\quad 
A,B\in\mathscr{B}(\mathscr{H}_d)\,.
\end{equation}
Suppose
\begin{equation}
\label{HS-basis}
\mathfrak{B}:=
\big\{\Gamma_\gamma\big\}%
_{\gamma=1}^{d^2}\,,\qquad
\lgroup \Gamma_{\gamma'},\Gamma_\gamma\,\rgroup_\textsc{hs}=
\delta_{\gamma,\gamma'}\,,
\end{equation}
is an orthonormal basis of $\mathscr{B}(\mathscr{H}_d)$ \cite{Schwinger60,Kimura03,Byrd03}, where 
${\delta_{\gamma,\gamma'}}$ is the Kronecker delta function.
Now we can resolve every operator ${A\in\mathscr{B}(\mathscr{H}_d)}$
in the basis $\mathfrak{B}$ as \cite{Fano57}
\begin{equation}
\label{A-resolution}
A=\sum_{\gamma=1}^{d^2}
\texttt{a}_\gamma\,\Gamma_\gamma\,,
\quad\mbox{where}\quad 
\texttt{a}_\gamma=\lgroup \Gamma_\gamma,A\,\rgroup_\textsc{hs}
\end{equation}
are complex numbers.
In this way, we also have the resolution of 
\begin{equation}
\label{rho-resolution}
\rho=\sum_{\gamma=1}^{d^2}
\texttt{r}_\gamma\,\Gamma_\gamma\,,
\quad\mbox{where}\quad 
\texttt{r}_\gamma=\lgroup \Gamma_\gamma,\rho\,\rgroup_\textsc{hs}\,.
\end{equation}

Born introduced the rule \cite{Born26} (see also \cite{von-Neumann55})
\begin{eqnarray}
\label{Born rule-1}
\langle A\rangle_{\rho}&=&
\text{tr}(\rho\, A)=\lgroup A^\dagger,\rho\rgroup_\textsc{hs}\\
\label{Born rule-2}
&=&\lgroup\rho,A\rgroup_\textsc{hs}
\end{eqnarray}
to calculate the average value of an operator $A$ by taking the statistical operator $\rho$.
Definition~\eqref{HS-inner-pro} of the inner product
is exploited to reach the last term in~\eqref{Born rule-1}, and through Hermiticity~\eqref{Herm-rho}, we get~\eqref{Born rule-2}.
By the rule, \eqref{Born rule-1}, one can realize 
\begin{equation}
\label{r}
\texttt{r}_\gamma= 
\langle \Gamma_\gamma^{\,\dagger} \rangle_\rho
=\overline{\langle \Gamma_\gamma \rangle}_\rho\,,
\end{equation} 
where the last equality is due to the conjugate symmetry
${\lgroup \Gamma,\rho\rgroup_\textsc{hs}=
	\overline{\lgroup \rho,\Gamma\rgroup}_\textsc{hs}}$
and \eqref{Born rule-2}.
The overline designates the complex conjugation.
The set of equations ${\langle \Gamma_\gamma^{\,\dagger} \rangle=\overline{\langle \Gamma_\gamma \rangle}}$ for every $\gamma$,
or ${\langle A^{\dagger} \rangle=\overline{\langle A \rangle}}$ 
for every ${A\in\mathscr{B}(\mathscr{H}_d)}$,
is equivalent to Hermiticity~\eqref{Herm-rho} of $\rho$.

Using \eqref{A-resolution} and \eqref{rho-resolution}, we can express \eqref{Born rule-2} as 
the standard inner product
\begin{equation}
\label{<A>-1}
\langle A\rangle_\rho
=\sum_{\gamma=1}^{d^2}\,
\overline{\texttt{r}}_{\gamma}\;\texttt{a}_\gamma
=\texttt{R}^\dagger\texttt{A}
\end{equation}
between 
${\texttt{R}:=(\texttt{r}_1,\cdots,\texttt{r}_{d^2})^\intercal}$
and
${\texttt{A}:=(\texttt{a}_1,\cdots,\texttt{a}_{d^2})^\intercal}$ \cite{Fano57,von-Neumann27}, where $\intercal$ stands for the transpose.
The column vectors ${\texttt{A},\texttt{R}\in\mathbb{C}^{d^2}}$
are the numerical representations of ${A,\rho\in\mathscr{B}(\mathscr{H}_d)}$
in basis~\eqref{HS-basis}, whereas expectation value \eqref{<A>-1} does not depend on the basis $\mathfrak{B}$ \cite{unitary-eq}.

Suppose ${A=A^\dagger}$ (depicts an observable) is a Hermitian operator, and  
\begin{equation}
\label{A}
A=\sum_{l=1}^{d}
a_l\,
|a_l\rangle\langle a_l|
\end{equation}
is its spectral decomposition.
Its expectation value [via \eqref{Born rule-1}]
\begin{equation}
\label{<A>-pro}
\langle A\rangle_\rho=
\sum_{l=1}^{d}a_l\,p_l\,,
\qquad
p_l=
\big\langle\,|a_l\rangle\langle a_l|\,\big\rangle_\rho\,,
\end{equation}
can be estimated by performing measurements in its eigenbasis 
${\{|a_l\rangle\}_{l=1}^{d}}$.
$p_l$ is the probability of getting the outcome, eigenvalue, $a_l$.
Due to \eqref{norm-rho} and \eqref{pos-rho}, one can realize
\begin{eqnarray}
\label{p-const1}
1&=&\sum_{l=1}^{d}
p_l=\text{tr}(\rho)\quad \mbox{and} \\  
\label{p-const2}
0&\leq&p_l=
\langle a_l|\,\rho\,|a_l\rangle
\quad \mbox{for all} \quad 
1\leq l\leq d\,.
\end{eqnarray}
In \eqref{p-const1}, the completeness relation
${\textstyle\sum\nolimits_{l=1}^{d}|a_l\rangle\langle a_l|=I}$ plays a role, where $I$ is the identity operator.	
The set of all probability vectors 
${\vec{p}:=(p_1,\cdots,p_d)}$
constitutes a probability space $\Omega_a$, that is---defined by
\eqref{p-const1} and \eqref{p-const2}---the standard ${(d-1)}$-simplex in the $d$-dimensional real vector space $\mathbb{R}^d$ \cite{Bengtsson06,Sehrawat17}.
One can perceive ${\langle A\rangle_\rho}$ in \eqref{<A>-pro} as a linear function from $\Omega_a$ into $\mathbb{R}$ and then can recognize
\begin{equation}
\label{<A>in}
\langle A\rangle\in[a_\text{min}\,,\,a_\text{max}]\,,
\end{equation}
where endpoints of the interval are the smallest $a_\text{min}$
and the largest $a_\text{max}$ eigenvalues of $A$.

Every classical (discrete) probability distribution also follows 
\eqref{p-const1} and \eqref{p-const2} \cite{Bengtsson06}.
The QCs become evident when we take two or more \emph{incompatible} observables (measurements), see below.
It is one of the most striking features of quantum physics that---has no classical analog---physically distinct measurements do exist, and one cannot estimate all the expectation values listed in $\overline{\texttt{R}}$ in \eqref{expt-values} by using a single setting for projective measurements \cite{SIC-POVM}. 
One requires at least ${d+1}$ settings.
Moreover, two measurement settings can be so different that if one always gets a definite outcome in one setting, (s)he can get totally random results in the other setting \cite{Ivanovic81,Wootters89}.
Such settings correspond to \emph{complementary} operators \cite{Schwinger60,Kraus87} that are building blocks of the unitary-basis presented in Sec.~\ref{sec:unitary-basis}.

Now let us take \textsc{n} number of operators: ${A,B,\cdots,C}$.
We can build a single matrix equation
\begin{equation}
\label{expt-values}
\underbrace{\begin{pmatrix}
	\langle A\rangle_\rho\\
	\langle B\rangle_\rho\\
	\vdots\\
	\langle C\rangle_\rho\\
	\end{pmatrix}}_{\displaystyle\textsf{E}}
= 
\underbrace{\begin{pmatrix}
	\texttt{a}_{\scriptscriptstyle 1} & 
	\texttt{a}_{\scriptscriptstyle 2} & 
	\cdots &
	\texttt{a}_{\scriptscriptstyle d^2}  \\
	\texttt{b}_{\scriptscriptstyle 1} & 
	\texttt{b}_{\scriptscriptstyle 2} & 
	\cdots &
	\texttt{b}_{\scriptscriptstyle d^2}  \\
	\vdots  & \vdots  & \ddots & \vdots  \\
	\texttt{c}_{\scriptscriptstyle 1} & 
	\texttt{c}_{\scriptscriptstyle 2} & 
	\cdots &
	\texttt{c}_{\scriptscriptstyle d^2}  \\
	\end{pmatrix}}_{\displaystyle \textbf{M} }
\underbrace{\begin{pmatrix}
	\overline{\texttt{r}}_{\scriptscriptstyle 1} \\
	\overline{\texttt{r}}_{\scriptscriptstyle 2} \\
	\vdots\\
	\overline{\texttt{r}}_{\scriptscriptstyle d^2} \\
	\end{pmatrix}}_{\displaystyle\overline{\texttt{R}}}
\end{equation} 
by combining equations such as \eqref{<A>-1}.
Equation \eqref{expt-values}
is nothing but the numerical representation of Born's rule~\eqref{Born rule-2} in basis~\eqref{HS-basis}.

We present this article by keeping the experimental scenario,
\begin{equation}
\label{expt-situ}
\parbox{0.85\columnwidth}
{
	a finite number of independent qudits are identically prepared in a quantum state $\rho$, and then individual qudits are measured using different settings for ${A,B,\cdots,C}$,
}
\end{equation}
in mind, where every expectation value is drawn from a same $\rho$.
Thus the subscript $\rho$ is omitted from 
$\langle \ \rangle_{\rho}$ at some places for simplicity of notation.
In other experimental situations---(\textit{i}) where one wants to entangle the qudit of interest to an ancillary system and then wants to perform a joint measurement or (\textit{ii}) where one desires to execute sequential measurements on the same qudit \cite{Busch14}---one can also adopt the above formalism.
There one may need to keep track of how the initial qudit's state gets transformed after an entangling operation or a measurement.
At each stage of an experiment, a $\rho$ must respect \eqref{Herm-rho}, \eqref{norm-rho}, and \eqref{0<=S_n}, and the mean values can be obtained by \eqref{expt-values}.

Matrix equation~\eqref{expt-values} has three parts $\displaystyle\overline{\texttt{R}}$, \textbf{M}, and \textsf{E}:
\begin{itemize}
	\item 
	Conditions~\eqref{Herm-rho}, \eqref{norm-rho}, and \eqref{0<=S_n} on a density operator $\rho$ enter through $\displaystyle\overline{\texttt{R}}$ and emerge as the QCs on the expectation values listed in \textsf{E}.
	In experiment situation~\eqref{expt-situ}, all the knowledge about state preparation goes into the column $\displaystyle\overline{\texttt{R}}$.
	\item
	From top to bottom, rows in the ${\textsc{n}\times d^{\,2}}$ matrix
	\textbf{M} completely specify ${A,B,\cdots,C}$. So \textbf{M} holds all, and only, the information about measurement settings. 	
	\item
	Conditions \eqref{Herm-rho}, \eqref{norm-rho}, and \eqref{0<=S_n} as well as the mean values in \textsf{E} do not depend on the choice of basis \cite{unitary-eq}.
	Therefore, the QCs on $\langle A\rangle,\langle B\rangle,\cdots,\langle C\rangle$ will be independent of the basis $\mathfrak{B}$.
	So one can adopt any basis that suits him or her best. 
	A basis only facilitates the transfer of constraints from a
	quantum state $\rho$ onto the expectation values in \textsf{E}.
\end{itemize}

Basically, one can achieve the QCs via a two-step procedure: 
\begin{enumerate}
	\item We need to express
	conditions \eqref{Herm-rho}, \eqref{norm-rho}, and \eqref{0<=S_n} for ${2\leq n\leq d}$
	in terms of
	${\{\overline{\texttt{r}}_\gamma\}_{\gamma=1}^{d^2}}$. 
	This delivers the QCs on mean values~\eqref{r} of the basis elements.
	\item 
	Then, we acquire the QCs on 
	$\langle A\rangle,\langle B\rangle,\cdots,\langle C\rangle$
	by matrix equation~\eqref{expt-values}.
\end{enumerate}
Let us focus on Step~1.
We already have condition~\eqref{Herm-rho}
in terms of
${\langle\Gamma_\gamma\rangle_\rho}$, see \eqref{r}.
To write the remaining conditions \eqref{norm-rho} and \eqref{0<=S_n} for ${2\leq n\leq d}$ in 
${\langle\Gamma_\gamma\rangle_\rho}$
terms, we need to compute 
\begin{equation}
\label{tr(rho^m)}
\text{tr}(\rho^m)=
\sum_{\gamma_1}\cdots
\sum_{\gamma_m}\,
\texttt{r}_{\gamma_1}\cdots\texttt{r}_{\gamma_m}
\text{tr}\left(\,\Gamma_{\gamma_1}\cdots\Gamma_{\gamma_m}\right)
\end{equation}
for every ${1\leq m\leq d}$.
One can view ${\text{tr}(\rho^m)}$ as a homogeneous polynomial of degree $m$, where average values~\eqref{r} are variables, and the constants 
${\text{tr}\left(\,\Gamma_{\gamma_1}\cdots\Gamma_{\gamma_m}\right)}$ are determined by basis~\eqref{HS-basis} only.
Hence $S_n$ of \eqref{S_n} is a $n$-degree polynomial, and
${0\leq S_n}$ [see \eqref{0<=S_n}] leads to a $n$-degree QC.

In \cite{Kimura03,Byrd03}, generators of the special unitary group
$SU(d)$---that with the identity operator compose an orthogonal basis of $\mathscr{B}(\mathscr{H}_d)$---are taken, and $\text{tr}(\rho^m)$ is obtained by using the Lie algebra of $SU(d)$.
The generators are ${d^{\,2}-1}$ traceless Hermitian operators, thus we call this basis the \emph{Hermitian}-basis [for ${d=2,3}$, see Appendix~\ref{sec:qubit} and Sec.~\ref{sec:qutrit}].
If all the \textsc{n} operators $A,B,\cdots,C$ are Hermitian operators, then it is better
to choose a Hermitian-basis because every number in \eqref{expt-values} will be a real number.
Since the state space
\begin{equation}
\label{set-of-states}
\mathcal{S}=\big\{\rho\in\mathscr{B}(\mathscr{H}_d)\ |\ 
\rho\text{ obeys \eqref{Herm-rho}, \eqref{norm-rho}, and 
	\eqref{0<=S_n}} \big\}
\end{equation}
is a compact and convex set \cite{Bengtsson06}, the corresponding collection 
of ${\texttt{R}=\overline{\texttt{R}}}$
forms a compact and convex set in $\mathbb{R}^{d^2-1}$ as the mapping 
${\rho\leftrightarrow\texttt{R}}$ is a homeomorphisms \cite{Rudin91}.
Every qudit's state $\rho$ is completely specified by ${d^{\,2}-1}$ real numbers in ${\texttt{R}}$ \cite{Kimura03}, where one of its components is fixed by normalization condition~\eqref{norm-rho}, that is,  $\textstyle\sum\nolimits_{\gamma=1}^{d^2}
\texttt{r}_\gamma\,\text{tr}(\Gamma_\gamma)=1$.

\begin{figure}
	\centering
	\includegraphics[width=0.3\textwidth]{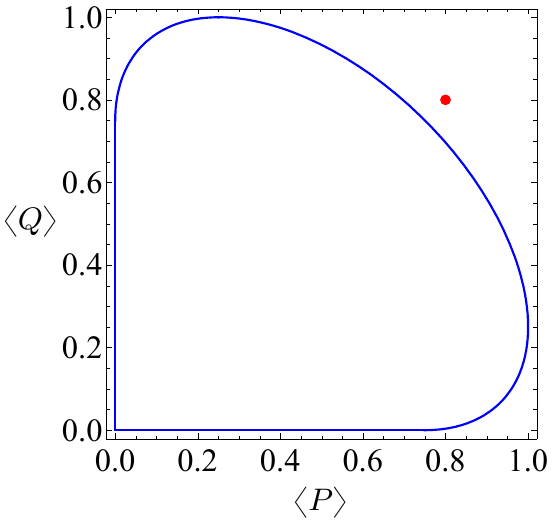}
	\caption{The permitted region $\mathcal{E}$---of the expectation values of two projectors
		$P$ and $Q$ described by the matrices in \eqref{PQ-mat}---is bounded by the (blue) closed-curve. $\mathcal{E}$ is the convex hull of ${(0,0)}$ and the ellipse obtained by \eqref{ellipse-PQ} and \eqref{G-PQ} with
		${|\langle a|b\rangle|^2=\text{tr}(PQ)=\tfrac{169}{675}}$.
		Clearly, the (red) point ${(0.8,0.8)}$ does not belong to the allowed region.
	In this example, hyperrectangle~\eqref{hyperrectangle} is the square ${[0,1]^{\times 2}}$.}
	\label{fig:region-2-pojs} 
\end{figure}

Next one can 
view \eqref{expt-values} as a linear transformation from $\mathbb{R}^{d^2-1}$ to $\mathbb{R}^{\textsc{n}}$.
Such a transformation is always continuous, and it maps a compact and convex set in $\mathbb{R}^{d^2-1}$ to a compact and convex set in $\mathbb{R}^{\textsc{n}}$
\cite{Rudin76,con-to-con}.
Therefore, for \textsc{n} observables (Hermitian operators), the set of expectation values
\begin{equation}
\label{set-of-expt}
\mathcal{E}:=\big\{\,\textsf{E}\ |\
\rho\in\mathcal{S}\, \big\}
\end{equation}
will be a \emph{compact} and \emph{convex} set 
[for example, see Figs. \ref{fig:region-2-pojs}, \ref{fig:regions-MUBs}, \ref{fig:regions}, \ref{fig:ellip-SUR}, and \ref{fig:regions-SIC-POVM}] in 
a hyperrectangle 
\begin{equation}
\label{hyperrectangle}
\mathcal{H}:=
[a_\text{min}, a_\text{max}]\times
[b_\text{min}, b_\text{max}]\times\cdots\times
[c_\text{min}, c_\text{max}]\subset\mathbb{R}^{\textsc{n}}
\end{equation}
described by the Cartesian product of the closed intervals, 
whose endpoints are the minimum and maximum eigenvalues of the operators.
$\mathcal{E}$ is also known as the
quantum convex support \cite{Weis11}.
Furthermore, each extreme point of $\mathcal{E}$ corresponds to a pure state that is an extreme point of $\mathcal{S}$.
Note that
Eq.~\eqref{expt-values} does (map $\mathcal{S}$ \emph{onto} $\mathcal{E}$ via $\rho\leftrightarrow\texttt{R}\rightarrow \textsf{E}$) not provide
a one-to-one correspondence between the state space $\mathcal{S}$ and  $\mathcal{E}$ unless there are $d^{\,2}$ linearly independent operators in the set ${\{A,B,\cdots,C,I\}}$.

In summary, $\mathcal{S}$ is an abstract set, we observe its \emph{image} $\mathcal{E}$
through an experiment scheme such as \eqref{expt-situ}.
The QCs---originate from \eqref{Herm-rho}, \eqref{norm-rho}, and \eqref{0<=S_n} via matrix equation~\eqref{expt-values}---bound the region $\mathcal{E}$. 
As the QCs are necessary and sufficient restrictions on the expectation values, any point outside $\mathcal{E}$ does not come from a quantum state, whereas every point in $\mathcal{E}$ corresponds to at least one quantum state. So as a whole $\mathcal{E}$ is the only \emph{allowed} region 
in the space of expectation values. Obviously,
one cannot achieve a region smaller than $\mathcal{E}$ without sacrificing a subset of quantum states.

Now we present 
all the above material by taking a \emph{standard} operator-basis.
With an orthonormal basis $\mathcal{B}$ of the Hilbert space $\mathscr{H}_d$, where
\begin{eqnarray}
\label{B_i}
\mathcal{B}&:=&
\big\{|j\rangle\,:\,j\in\mathbb{Z}_d\big\}\,,\\
\label{Z_d}
\mathbb{Z}_d&:=&\{\,j\,\}_{j=0}^{d-1}\,,
\qquad\mbox{and}\qquad\\
\label{orth-|j>}
\text{tr}\big(|j\rangle\langle k|\big)&=&
\langle k|j\rangle=\delta_{j,k}\,,
\end{eqnarray}
one can construct the standard operator-basis
\begin{equation}
\label{Stnd-Basis}
\mathfrak{B}_\text{st}:=\big\{|j\rangle\langle k|\,:\,j,k\in\mathbb{Z}_d\big\}
\end{equation}
of $\mathscr{B}(\mathscr{H}_d)$.
Instead of a single index $\gamma$ that runs from $1$ to ${d^{\,2}}$,
here we have two indices $j$ and $k$ for a basis element, each of them runs from $0$ to ${d-1}$. The orthonormality condition 
\begin{equation}
\label{orth-|j><k|}
\big\lgroup\, |j'\rangle\langle k'|\,,\,|j\rangle\langle k|\,\big\rgroup_\textsc{hs}
=\langle j'|j\rangle\langle k|k'\rangle
=\delta_{j,j'}\,\delta_{k,k'}
\end{equation} 
for $\mathfrak{B}_\text{st}$ is ensured by orthonormality relation \eqref{orth-|j>} of $\mathcal{B}$.
In basis \eqref{Stnd-Basis}, the resolution of an operator $A$ and of a qudit's state $\rho$ are
\begin{eqnarray}
\label{A-in-stnd-basis}
A&=&
\sum_{j,k\,\in\,\mathbb{Z}_d}
\texttt{a}_{jk}\,|j\rangle\langle k|
\quad\mbox{with}\quad 
\texttt{a}_{jk}=\langle j|\,A\,|k\rangle\quad\mbox{and}\qquad\\
\label{rho-in-stnd-basis}
\rho&=&
\sum_{j,k\,\in\,\mathbb{Z}_d}
\texttt{r}_{jk}\,|j\rangle\langle k|
\quad\mbox{with}\quad 
\texttt{r}_{jk}=\langle j|\,\rho\,|k\rangle\,,
\end{eqnarray}
respectively.
The above coefficients $\texttt{a}$ and $\texttt{r}$ are obtained through \eqref{A-resolution} and \eqref{rho-resolution}, correspondingly.

Numerical representation~\eqref{<A>-1} of Born's rule now
becomes
\begin{equation}
\label{<A>-in-stnd-basis-2}
\langle A\rangle
=
\sum_{j,k}\,
\overline{\texttt{r}}_{jk}\,\texttt{a}_{jk}\\
=
\sum_{j,k}\,
\texttt{r}_{kj}\,\texttt{a}_{jk}\,,
\end{equation}
where the second equality is due to the Hermiticity:
\begin{equation}
\label{r_kj}
\texttt{r}_{jk}=
\big\langle\, |k\rangle\langle j|\,  \big\rangle_\rho=
\overline{\texttt{r}}_{kj}
\quad\mbox{for all}\quad
j,k\in\mathbb{Z}_d
\end{equation}
is a manifestation of \eqref{r}.
In standard basis~\eqref{Stnd-Basis}, matrix equation~\eqref{expt-values} transpires as
\begin{equation}
\label{expt-values-std}
\underbrace{\begin{pmatrix}
	\langle A\rangle_{\rho}\\
	\langle B\rangle_{\rho}\\
	\vdots\\
	\langle C\rangle_{\rho}
	\end{pmatrix}}_{\displaystyle\textsf{E}}
= 
\underbrace{\begin{pmatrix}
	\texttt{a}_{\scriptscriptstyle 0,0} & 
	\texttt{a}_{\scriptscriptstyle 0,1} & \cdots &
	\texttt{a}_{\scriptscriptstyle d-1,d-1}  \\
	\texttt{b}_{\scriptscriptstyle 0,0} & 
	\texttt{b}_{\scriptscriptstyle 0,1} & \cdots &
	\texttt{b}_{\scriptscriptstyle d-1,d-1}  \\
	\vdots  & \vdots  & \ddots & \vdots  \\
	\texttt{c}_{\scriptscriptstyle 0,0} & 
	\texttt{c}_{\scriptscriptstyle 0,1} & \cdots &
	\texttt{c}_{\scriptscriptstyle d-1,d-1}  \\   
	\end{pmatrix}}_{\displaystyle \textbf{M} }
\underbrace{\begin{pmatrix}
	\overline{\texttt{r}}_{\scriptscriptstyle 0,0} \\
	\overline{\texttt{r}}_{\scriptscriptstyle 0,1}\\
	\vdots\\
	\overline{\texttt{r}}_{\scriptscriptstyle d-1,d-1}
	\end{pmatrix}}_{\overline{\displaystyle\texttt{R}}_d}.\qquad
\end{equation}

Next, to express conditions \eqref{norm-rho} and \eqref{0<=S_n} for ${2\leq n\leq d}$ in 
${\texttt{r}_{jk}}$
terms, we need to represent
$\text{tr}(\rho^m)$ for every ${1\leq m\leq d}$ as a function of
$\{\texttt{r}_{jk}\}$.
Orthonormality relation~\eqref{orth-|j>} also yields the rule for composition
\begin{equation}
\label{comp}
|j'\rangle\langle k'|\,|j\rangle\langle k|=
\delta_{j,k'}\,|j'\rangle\langle k|\,,
\end{equation}
which gives rise to matrix multiplication
in the matrix mechanics \cite{Heisenberg25,Born25,Born25-b,Dirac25,Waerden68}.
Particularly here it is very easy to obtain
\begin{equation}
\label{rho^m-stnd}
\rho^m=
\sum_{j_1}\cdots
\sum_{j_{m+1}}
\texttt{r}_{j_1j_2}\,\texttt{r}_{j_2j_3}\cdots\texttt{r}_{j_mj_{m+1}}
\,|j_1\rangle\langle j_{m+1}|\,.\\
\end{equation}
Then, through \eqref{orth-|j>} and the linearity of trace, we secure
\begin{equation}
\label{tr(rho^m)-stnd}
\text{tr}(\rho^m)=
\sum_{j_1}\cdots
\sum_{j_m}\,
\texttt{r}_{j_1j_2}\,\texttt{r}_{j_2j_3}\cdots\texttt{r}_{j_mj_{1}}\,.
\end{equation}
One can compare \eqref{tr(rho^m)-stnd} with its general form~\eqref{tr(rho^m)}.
Let us explicitly write conditions \eqref{norm-rho} and \eqref{0<=S_n} for ${n=2,3,4}$ \cite{Kimura03,Fano57}:
\begin{eqnarray}
\label{1=S_1}
\sum_{j}
\texttt{r}_{jj}=\text{tr}(\rho)&=& 1\,,\\
 \label{0<=S_2}
\texttt{R}^\dagger \texttt{R}=
\sum_{jk}
 |\texttt{r}_{jk}|^2=\text{tr}(\rho^2)&\leq& 1\,,\\
 \label{0<=S_3}
 3\,\text{tr}(\rho^2)-2\,\text{tr}(\rho^3)&\leq& 1\,,\\
\label{0<=S_4}
6\,\text{tr}(\rho^2)-8\,\text{tr}(\rho^3)-3\,\big(\text{tr}(\rho^2)\big)^2+6\,\text{tr}(\rho^4)&\leq& 1\qquad
\end{eqnarray}
deliver linear, quadratic, cubic, and quartic QCs.
In \eqref{1=S_1} and \eqref{0<=S_2}, \eqref{tr(rho^m)-stnd}
and the column vector \texttt{R} from \eqref{expt-values-std} are used.

As a pure state ${\rho=\rho^2}$ is an extreme point of the state space
$\mathcal{S}$ [defined in \eqref{set-of-states}], it saturates inequalities \eqref{0<=S_n} for all ${n=2,\cdots,d}$ \cite{Byrd03}.
A pure state corresponds to a ket, and a qudit's ket can be parametrized by a set of ${2(d-1)}$ real numbers by ignoring an overall phase factor
(for example, see \cite{Arvind97}):
\begin{eqnarray}
\label{|psi>}
|\psi\rangle&=&
  |0\rangle \cos\theta_0+\nonumber\\
&&|1\rangle \sin\theta_0\cos\theta_1\,e^{\text{i}\phi_1}+\nonumber\\
&&|2\rangle \sin\theta_0\sin\theta_1\cos\theta_2\,e^{\text{i}\phi_2}+\nonumber\\
&&\qquad\cdots+\nonumber\\
&&|d-2\rangle \sin\theta_0\sin\theta_1\cdots\cos\theta_{d-2}\,e^{\text{i}\phi_{d-2}}+
\nonumber\\
&&|d-1\rangle \sin\theta_0\sin\theta_1\cdots\sin\theta_{d-2}\,e^{\text{i}\phi_{d-1}}\,,
\end{eqnarray}
where ${\text{i}=\sqrt{-1}}$, $\theta_l\in[0,\tfrac{\pi}{2}]$ for all ${l=0,\cdots,d-2}$, and $\phi_{l'}\in[0,2\pi)$ for every ${l'=1,\cdots,d-1}$.
Thus the pure state ${\rho_\text{pure}=|\psi\rangle\langle\psi|}$ and the corresponding column vector ${\texttt{R}^{(\text{pure})}_d}$ [see \eqref{expt-values-std} for its complex conjugate] are specified by the ${2(d-1)}$ real numbers \cite{Bengtsson06},
for instance, 
\begin{eqnarray}
\label{R-pure-2}
&\texttt{R}^{(\text{pure})}_2=\begin{pmatrix}
(\cos\theta_0)^2 \\
\cos\theta_0\sin\theta_0\,e^{-\text{i}\phi_1}\\
\cos\theta_0\sin\theta_0\,e^{\text{i}\phi_1}\\
(\sin\theta_0)^2
\end{pmatrix}&
 \mbox{and}\qquad\quad
\\
\label{R-pure-3}
&\texttt{R}^{(\text{pure})}_3=\begin{pmatrix}
(\cos\theta_0)^2 \\
\cos\theta_0\sin\theta_0\cos\theta_1\,e^{-\text{i}\phi_1}\\
\cos\theta_0\sin\theta_0\sin\theta_1\,e^{-\text{i}\phi_2}\\
\cos\theta_0\sin\theta_0\cos\theta_1\,e^{\text{i}\phi_1}\\
{(\sin\theta_0\cos\theta_1)}^2\\
(\sin\theta_0)^2\cos\theta_1\sin\theta_1\,e^{\text{i}(\phi_1-\phi_2)}\\
\cos\theta_0\sin\theta_0\sin\theta_1\,e^{\text{i}\phi_2}\\
(\sin\theta_0)^2\cos\theta_1\sin\theta_1\,e^{-\text{i}(\phi_1-\phi_2)}\\
{(\sin\theta_0\sin\theta_1)}^2
\end{pmatrix}.&
\end{eqnarray}
By plugging ${\texttt{R}^{(\text{pure})}_d}$ in Eq.~\eqref{expt-values-std}, one can reach all those points
in $\mathcal{E}$ [defined in \eqref{set-of-expt}] that correspond to pure states in $\mathcal{S}$.
All the extreme points of $\mathcal{E}$ will be a subset of these points.

In the following, we demonstrate a procedure to built a combined uncertainty  measure on $\mathcal{E}$ for Hermitian operators ${A,B,\cdots,C}$.
In the case of a non-Hermitian operator, considering \cite{Adagger}, one can talk about 
uncertainty measures for the two Hermitian operators
$A^{\scriptscriptstyle(+)}=\tfrac{1}{2}(A+A^\dagger)$ and $A^{\scriptscriptstyle(-)}=\tfrac{1}{2\text{i}}(A-A^\dagger)$. Note that $A^{\scriptscriptstyle(+)}$ and $A^{\scriptscriptstyle(-)}$ commutes if and only if $A$---is a normal operator---commutes with $A^\dagger$.

The standard deviation 
\begin{eqnarray}
\label{std-dev}
\Delta A&=&\sqrt{\langle A^2\rangle-\langle A\rangle^2}\nonumber\\
&=&\sqrt{\sum_{l=1}^{d}a_l^2\,p_l
	-\left(\sum_{l=1}^{d}a_l\,p_l\right)^2}
\end{eqnarray}
can be viewed---through the first equality---as a concave function on the allowed region for $\{A,A^2\}$, which is 
the convex hull of ${\{(a_l,a_l^2)\}_{l=1}^d}$, where $a_l$ is an eigenvalue of $A$ [see \eqref{A}].
By finding the absolute minimum of ${\Delta A+\Delta B+\cdots+\Delta C}$ on
the permitted region for $\{A,A^2,B,B^2,\cdots,C,C^2\}$, one can have a tight UR 
based on the standard deviations 
[for example, see \eqref{std-J 9}--\eqref{aq-std-J}].

If one wants to built a UR in the case of two projective measurements
described by ${\{|a_l\rangle\langle a_l|\}_{l=1}^d}$
and ${\{|b_k\rangle\langle b_k|\}_{k=1}^d}$, then one can consider the
permissible region of the two probability vectors ${\vec{p}=(p_1,\cdots,p_d)}$ and 
${\vec{q}=(q_1,\cdots,q_d)}$, where 
${p_l=\big\langle |a_l\rangle\langle a_l|\big\rangle_\rho}$ 
[see \eqref{<A>-pro}] and
${q_k=\big\langle |b_k\rangle\langle b_k|\big\rangle_\rho}$.
There are many uncertainty measures for $\vec{p}$ (and $\vec{q}\,$)---thanks to Shannon \cite{Shannon48}, R\'{e}nyi \cite{Renyi61}, and
Tsallis \cite{Tsallis88}---and many associated URs \cite{Coles17,Maassen88}.
Moreover, with the probability vector $\vec{p}$, we can calculate the expectation value of any function of the Hermitian operator ${A}$ [given in \eqref{A}] as well as its standard deviation \eqref{std-dev}.
Now suppose we have no access to the individual probabilities $p_l$, but only to the expectation value 
${\langle A\rangle}$, then we can construct uncertainty or certainty measures as follows.

Let us recall from \eqref{<A>in} that 
${\langle A\rangle\in[a_\text{min}\,,\,a_\text{max}]}$, and we are interested in the case ${a_\text{min}\neq\,a_\text{max}}$.
We call $\rho$ an \emph{eigenstate} corresponding to an eigenvalue $a$ of $A$
if and only if ${A\rho=a\rho=\rho A}$.
If ${\langle A\rangle_\rho=a_\text{min}}$ then we can say for sure: ($i$) qudits are prepared in a minimum-eigenvalue-state of ${A}$ and ($ii$) every outcome 
${a_l\neq a_\text{min}}$ will never occur in a future projective measurement
${\{|a_l\rangle\langle a_l|\}_{l=1}^d}$
for $A$.
So, only in the two cases ${\langle A\rangle_\rho=a_\text{min},a_\text{max}}$, we have a minimum possible uncertainty
about $\rho$ (if it is unknown) in which the individual qudits are identically prepared in \eqref{expt-situ}
and about the results of a future measurement for $A$.
Therefore, for an uncertainty measure, we require a continuous function on the interval $[a_\text{min}\,,\,a_\text{max}]$
that reaches its absolute minimum at both the endpoints.
Furthermore, mixing states, ${w\rho+(1-w)\rho'=\rho_{\text{mix}}}$ with ${0\leq w\leq1}$,
yields the convex sum ${w\langle A\rangle_\rho+(1-w)\langle A\rangle_{\rho'}=\langle A\rangle_{\rho_{\text{mix}}}}$, and it does not decrease uncertainty (or increase certainty).
A suitable concave (convex) function can be taken as a measure of uncertainty (certainty) because it does not decrease (increase) under such mixing.

The two positive semi-definite operators 
\begin{equation}
\label{Adot}
\dot{A}:=\frac{a_\text{max}\,I-A}{a_\text{max}-a_\text{min}}
\quad\text{and}\quad
\mathring{A}:=\frac{A-a_\text{min}\,I}{a_\text{max}-a_\text{min}}\,,
\end{equation}
are such that ${\dot{A}+\mathring{A}}$ is the identity operator ${I\in\mathscr{B}(\mathscr{H}_d)}$, and we only need ${\langle A\rangle}$ to compute both
${\langle \dot{A}\rangle,\langle \mathring{A}\rangle\in[0,1]}$.
Now we can define concave and convex functions of ${\langle A\rangle}$
that fulfill the above requirements:
\begin{eqnarray}
\label{H(A)}
H(\langle A\rangle)&=&-(\langle \dot{A}\rangle\ln\,\langle \dot{A}\rangle+\langle \mathring{A}\rangle\ln\,\langle \mathring{A}\rangle)\,,
\\
\label{u(A)}
u_\kappa(\langle A\rangle)&=&{\langle \dot{A}\rangle}^\kappa+{\langle \mathring{A}\rangle}^\kappa\,,\quad 0<\kappa<\infty\,,
\quad\text{and}\qquad\\
\label{umax(A)}
u_\text{max}(\langle A\rangle)&=&\max\,\{\,
\langle \dot{A}\rangle\,,\,\langle \mathring{A}\rangle\,\}\,.
\end{eqnarray}
One can easily show that $H$ and $u_\kappa$ for all ${0<\kappa<1}$ are concave functions, whereas $u_\kappa$ for all ${1<\kappa<\infty}$ and $u_\text{max}$ are convex functions.
For ${\kappa=1}$, ${u_\kappa(\langle A\rangle)=1}$ for every $\langle A\rangle$, and thus it is neither a genuine measure of uncertainty nor of certainty.

With $u_\kappa$ one can create quantities like R\'{e}nyi's and
Tsallis' entropies, and $H$ of \eqref{H(A)} is like the Shannon entropy 
but, in general, it is different from ${-\sum_{l=1}^{d}p_l\ln p_l}$. 
If $A$ only has two distinct eigenvalues, then $\dot{A}$ and $\mathring{A}$ become
mutually orthogonal projectors, and
\eqref{H(A)} turns into the standard form of Shannon entropy [for example, see Appendix~\ref{sec:qubit}].
Note that Shannon's and Tsallis' entropies are concave functions but not all 
R\'{e}nyi's entropies are.

The ranges of the above functions are ${H\in[0,\ln2]}$, ${u_\kappa\in[1,2^{1-\kappa}]}$ for ${0<\kappa<1}$
and ${u_\kappa\in[2^{1-\kappa},1]}$ for ${1<\kappa<\infty}$, and
${u_\text{max}\in[\tfrac{1}{2},1]}$.
As desired, all the above concave (convex) functions reach their absolute minimum (maximum) when
${\langle A\rangle=a_\text{min},a_\text{max}}$. 
In the case of a non-degenerate eigenvalue $a_\text{min}$, we will be even more certain that there is only one (pure) eigenstate state 
${|a_\text{min}\rangle\langle a_\text{min}|}$ that can provide
${\langle A\rangle=a_\text{min}}$, and similarly for a non-degenerate $a_\text{max}$.
Like the standard deviation \eqref{std-dev},
all the concave (convex) functions in \eqref{H(A)}--\eqref{umax(A)} 
attain their absolute maximum (minimum) when
${\langle A\rangle=\tfrac{1}{2}(a_\text{min}+a_\text{max})}$.
Both a ket ${\tfrac{1}{\sqrt{2}}(|a_\text{min}\rangle+e^{\text{i}\phi}|a_\text{max}\rangle)}$, ${\phi}$ is a real number, and a state that is the equal mixture of $|a_\text{min}\rangle\langle a_\text{min}|$ and $|a_\text{max}\rangle\langle a_\text{max}|$ provide the expectation value ${\langle A\rangle=\tfrac{1}{2}(a_\text{min}+a_\text{max})}$.
Since the equal superposition ket gives the maximum standard deviation of $A$, the ket plays an important role in the quantum metrology \cite{Giovannetti06}
and to determine a fundamental limit on the speed of unitary evolution generated by $A$ \cite{Mandelstam45,Margolus98,Levitin09}.

The sum of concave functions is a concave function, for example, 
\begin{equation}
\label{Habc}
H(\textsf{E}):=
H(\langle A\rangle)+H(\langle B\rangle)+\cdots+H(\langle C\rangle)\,,
\end{equation}
where every $H$ is defined according to \eqref{H(A)}.
One can view \eqref{Habc} as a measure of combined uncertainty on the allowed region $\mathcal{E}$. 
Its global minimum, say, $\mathfrak{h}$
will occur at the extreme points of $\mathcal{E}$ (see Theorem~${3.4.7}$ and Appendix~A.3 in \cite{Niculescu93}).
As every extreme point of $\mathcal{E}$ is related to a
pure state, one can find the minimum by changing at most ${2(d-1)}$ parameters that
appear in \eqref{|psi>} and then can enjoy the \emph{tight} UR
${\mathfrak{h}\leq H(\textsf{E})}$.
If a vertex of hyperrectangle~\eqref{hyperrectangle} is a part of $\mathcal{E}$
only then the lower bound $\mathfrak{h}$ becomes (trivial) 0.
It only happens when there exists a ket ${|\text{e}\rangle}$ that is a maximum- or 
minimum-eigenvalue-ket of every operator in ${\{A,B,\cdots,C\}}$.
There are examples in \cite{Sehrawat18} where all ${A,B,\cdots,C}$ share a common eigenket, thus usual URs---based on probabilities associated with projective measurements for ${A,B,\cdots,C}$ or based on the standard deviations ${\Delta A,\Delta B,\cdots,\Delta C}$---become trivial while ${0<\mathfrak{h}\leq H(\textsf{E})}$.
Like \eqref{Habc}, one can built combined uncertainty or certainty measures (and relations) by picking concave or convex functions from \eqref{u(A)} and \eqref{umax(A)}.
If one chooses 
a measure that is neither a concave nor convex function then its absolute extremum
can occur inside $\mathcal{E}$.
The above technique is applied to derive tight URs and CRs in \cite{Riccardi17,Sehrawat17} and in the subsequent sections.

Apart from a few exceptions, it is not clear to us whether we can interpret a QC as a bound on a combined uncertainty or certainty.
On the other hand, a UR puts a lower limit on a combined uncertainty, and it can also be perceived as
a constraint on mean values as every uncertainty measure is (not necessarily concave or convex but) their function.
Suppose we identify a region in hyperrectangle~\eqref{hyperrectangle} with a UR, for example,
\begin{equation}
\label{R_H}
\mathcal{R}_H:=
\big\{ (\textbf{a},\cdots,\textbf{c})\in\mathcal{H}
\ |\ 
{\mathfrak{h}\leq H(\textbf{a})+\cdots+H(\textbf{c})}
\big\}\,,
\end{equation}
where $H(\textbf{a})$ is obtained by replacing ${\langle A\rangle}$
with \textbf{a} in ${\langle \dot{A}\rangle}$, ${\langle \mathring{A}\rangle}$, and
then in \eqref{H(A)}; likewise, ${H(\textbf{c})}$ has the same functional form as 
$H(\langle C\rangle)$.
One can easily prove that $\mathcal{R}_H$ is a convex set. 
Obviously, ${\mathcal{E}}$ will be contained in $\mathcal{R}_H$, 
there will be no $\rho$ for ${(\textbf{a},\cdots,\textbf{c})\in\mathcal{R}\setminus\mathcal{E}}$ such that
${(\textbf{a},\cdots,\textbf{c})=(\langle A\rangle_{\rho},\cdots,\langle C\rangle_{\rho})}$
holds, and such points cannot be realized experimentally in scheme~\eqref{expt-situ}.
The relative complement of $\mathcal{E}$ in $\mathcal{R}$ is denoted by $\mathcal{R}\setminus\mathcal{E}$. 
One can also observe that
if ${(\textbf{a},\textbf{b},\cdots,\textbf{c})}$ belongs to $\mathcal{R}_H$
then ${(\textbf{a}',\textbf{b},\cdots,\textbf{c})}$, where ${\textbf{a}'=a_{\text{min}}+a_{\text{max}}-\textbf{a}}$, will also belong to $\mathcal{R}_H$ because ${H(\textbf{a})=H(\textbf{a}')}$.
In the case of ${\textbf{a}'\neq\textbf{a}}$, only one of the two points 
can be allowed, because a single quantum state cannot provide two different expectation values of $A$.
By taking a few examples in this paper, 
the gap $\mathcal{R}\setminus\mathcal{E}$ between the two regions is exhibited in Figs. \ref{fig:region-2-pojs}, \ref{fig:regions-MUBs}, \ref{fig:regions}, \ref{fig:ellip-SUR}, and \ref{fig:regions-SIC-POVM}.

If we have to provide a yes/no answer to a question such as: 
can ${0.8}$ and ${0.8}$ be the expectation values ${\langle P\rangle_\rho}$ 
and ${\langle Q\rangle_\rho}$, where $P$ and $Q$ are rank-1 projectors represented by
\begin{equation}
\label{PQ-mat}
\begin{pmatrix}
\frac{1}{75} & -\frac{\text{i}}{15} & \frac{7}{15}\\[0.4em]
\frac{\text{i}}{15} & \hphantom{-}\frac{1}{3} & \frac{7\,\text{i}}{15}\\[0.4em]
\frac{7}{75} & -\frac{7\,\text{i}}{15} & \frac{49}{75}
\end{pmatrix}
\quad\mbox{and}\quad
\frac{1}{9}\begin{pmatrix}
\hphantom{-}4 & \hphantom{-}2 & -4\\[0.4em]
\hphantom{-}2 & \hphantom{-}1 & -2\\[0.4em]
-4 & -2 & \hphantom{-}4
\end{pmatrix},
\end{equation}
respectively, in some orthonormal basis of $\mathscr{H}_3$?
Then, a \emph{clear} answer can be given with the allowed region. 
Suppose 
${P=|a\rangle\langle a|}$
and
${Q=|b\rangle\langle b|}$ are two rank-1 projectors on a $d$-dimensional Hilbert space $\mathscr{H}_d$
such that ${0<|\langle a|b\rangle|<1}$ (non-commuting).
For ${d=2}$, their allowed region $\mathcal{E}$
is determined by
\begin{eqnarray}
\label{ellipse-PQ}
&&
\textsf{E}^\intercal\, \textbf{G}^{-1}\textsf{E} \leq 1\,,
\quad \mbox{where} \quad
\textsf{E}=\begin{pmatrix}
2\,\langle P\rangle-1\\
2\,\langle Q\rangle-1
\end{pmatrix}
\ \mbox{and}\qquad
\\
&&
\label{G-PQ}
\textbf{G}= 
\begin{pmatrix}
1 & {\scriptstyle 2|\langle a|b\rangle|^2-1}\\
{\scriptstyle 2|\langle a|b\rangle|^2-1}& 1
\end{pmatrix}.
\qquad
\end{eqnarray}
One can see through \eqref{angle-bt-axis} that \eqref{ellipse-PQ} and \eqref{ellipse}
are the same for a qubit.
In the case of ${d>2}$, the allowed region will be the convex hull of the elliptic region specified by the inequality in \eqref{ellipse-PQ} and the point $(0,0)$
\cite{Lenard72}; see also \cite{Sehrawat17}.
This point is given by all those states that lie in the orthogonal complement of
${\{P,Q\}}$. 
These states are the common eigenstates of $P$ and $Q$.
By the way, a UR become a trivial statement in this case.

Answer to the above question is ``no'' because the point ${(0.8,0.8)}$ falls outside the allowed region as shown in Fig.~\ref{fig:region-2-pojs}.
If one asks a similar question for a set of commuting operators
${\{A,B,\cdots,C\}}$, then the permitted region will be the convex hull of 
$\{(\langle \text{e}_l|A|\text{e}_l\rangle,
\langle \text{e}_l|B|\text{e}_l\rangle,\cdots,
\langle \text{e}_l|C|\text{e}_l\rangle)\}_{l=1}^{d}$,
where ${\{|\text{e}_l\rangle \}_{l=1}^{d}}$
is their common eigenbasis.


\section{The unitary operator basis}\label{sec:unitary-basis}

With orthonormal basis \eqref{B_i}
of the Hilbert space $\mathscr{H}_d$,
we can built a pair of (complementary) unitary
operators
\begin{eqnarray}
\label{X_i}
X&:=&\sum_{j\,\in\, \mathbb{Z}_d}
|j+1\rangle\langle j| 
\qquad\quad\qquad(X^d=I)\quad\mbox{and}\qquad\\
\label{Z_i}
Z&:=&\sum_{j\,\in\, \mathbb{Z}_d}
\omega^{\,j}\,|j\rangle\langle j| 
\qquad\quad\qquad\  (Z^d=I)
\end{eqnarray}
thanks to Weyl \cite{Weyl32} and Schwinger~\cite{Schwinger60},
where ${j+1}$ is the modulo-$d$ addition,
${\omega=\exp(\text{i}\tfrac{2\pi}{d})}$, and
$\mathbb{Z}_d$ is defined in \eqref{Z_d}.
Under the operator multiplication, $X$ and $Z$ generate the discrete Heisenberg-Weyl group \cite{Weyl32,Durt10}.
The group members follow the Weyl commutation relation \cite{Weyl32}
\begin{equation}
\label{Weyl commutation}
Z^zX^x=
\omega^{\,xz}\, X^xZ^z \quad\text{for every}\quad x,z\in\mathbb{Z}_d\,,
\end{equation} 
and the property 
\begin{equation}
\label{traceless}
\text{tr}(X^xZ^z)= d\,\delta_{x,0}\delta_{z,0}\,.
\end{equation}

A subset of the Weyl group
\begin{equation}
\label{Unitary-Basis}
\mathfrak{B}_{\text{uni}}:=\big\{X^xZ^z:x,z\in\mathbb{Z}_d\big\}
\end{equation} 
forms an orthogonal basis of $\mathscr{B}(\mathscr{H}_d)$, where the orthogonality relation
\begin{equation}
\label{orth-XZ}
\big\lgroup X^{x'}Z^{z'},X^xZ^z\big\rgroup_\textsc{hs}=
\text{tr}\big(X^{x-x'}Z^{z-z'}\big)=
d\,\delta_{x,x'}\delta_{z,z'}\qquad
\end{equation}
is a consequence of \eqref{traceless} \cite{Schwinger60}.
All the elements in basis~\eqref{Unitary-Basis} are unitary operators and traceless [see \eqref{traceless}] except the identity operator that corresponds to ${x=0=z}$.
Basis~\eqref{Unitary-Basis} is called the \emph{unitary}-basis.

According to \eqref{rho-resolution} and \eqref{r}, a
statistical operator can be represented as
\begin{equation}
\label{rho-in-XZ}
\rho=\tfrac{1}{d}\sum_{x,z\,\in\,\mathbb{Z}_d}
\overline{\langle X^xZ^z\rangle}_\rho\ X^xZ^z
\end{equation}
in the basis $\mathfrak{B}_{\text{uni}}$. Here, the conditions for normalization \eqref{norm-rho} and for Hermiticity \eqref{r} become
${\langle X^0Z^0\rangle=1}$ and 
\begin{equation}
\label{<XxZz>}
\overline{\langle X^xZ^z\rangle}=\langle(X^xZ^z)^\dagger\rangle=
\omega^{\,xz}\langle X^{-x}Z^{-z}\rangle\,,
\end{equation}
respectively.
The second equality in \eqref{<XxZz>} is obtained by the virtue of \eqref{Weyl commutation}.
The inverse of a basis element, ${(X^xZ^z)^\dagger}$, 
does not always belong to basis~\eqref{Unitary-Basis} but
to the Weyl group.
Whereas both ${X^xZ^z}$ and ${X^{-x}Z^{-z}}$ are members of 
$\mathfrak{B}_{\text{uni}}$, and their mean values are related through~\eqref{<XxZz>} (in this regard, see also \cite{Adagger}).

Taking the general form, \eqref{tr(rho^m)},
one can easily express $\text{tr}(\rho^m)$ in the unitary-basis
by using \eqref{Weyl commutation}, \eqref{traceless}, and \eqref{<XxZz>},
for example, \begin{eqnarray}
\label{tr_rho2-exp}
\text{tr}(\rho^2)
&=&\tfrac{1}{d}\sum_{x,z}
{|\langle X^xZ^z \rangle|}^{\,2}\quad\mbox{and}\\
\label{tr_rho3-exp}
\text{tr}(\rho^3)&=&\tfrac{1}{d^2}
\sum_{x_1,z_1}\sum\limits_{x_2,z_2}
\langle X^{-x_1}Z^{-z_1} \rangle\,
\langle  X^{-x_2}Z^{-z_2} \rangle\times
\nonumber\\
&&\qquad
\langle X^{ x_1+x_2}Z^{ z_1+z_2} \rangle\;
\omega^{z_1(x_1+x_2)+z_2x_2}\,.\quad\quad
\end{eqnarray}
Then, one can draw QCs on the expectation values of the Weyl operators 
from~\eqref{0<=S_n}.

In the case of a prime dimensional $d$,
the basis $\mathfrak{B}_{\text{uni}}$---without the identity operator---can be divided into ${d+1}$ disjoint subsets 
\begin{eqnarray}
\label{d+1 subsets}
&&\big\{\mathcal{C}^{(1,z)}\,|\,z\in\mathbb{Z}_d\big\}
\cup\big\{\mathcal{C}^{(0,1)}\big\}
\,,\quad \mbox{where} \\
\label{Cxz}
&&\mathcal{C}^{(x,z)}:=
\big\{ X^{kx} Z^{kz}\,|\,
k\in\mathbb{Z}_d\ \ \mbox{and}\ \ k\neq0\big\}
\end{eqnarray} 
carries ${d-1}$ pairwise commuting operators \cite{Bandyopadhyay02,Englert01}.
Hence, one can find a common eigenbasis of the operators in $\mathcal{C}^{(x,z)}$.
In fact, there exists a complete set of ${d+1}$ MUBs
of $\mathscr{H}_{d}$ \cite{Ivanovic81,Wootters89,Bandyopadhyay02}:
\begin{equation}
\label{d+1 bases}
\big\{\mathcal{B}^{\scriptscriptstyle(z)}\,|\,z\in\mathbb{Z}_d\big\}
\cup\big\{\,\mathcal{B}\,\big\}
\end{equation}
are eigenbases for the subsets in 
\eqref{d+1 subsets}.
Our original basis $\mathcal{B}$ in \eqref{B_i} is an eigenbasis of ${Z\in\mathcal{C}^{(0,1)}}$ [see \eqref{Z_i}].
Let us define the remaining bases as \cite{Bandyopadhyay02,Englert01}
\begin{equation}
\label{B^z}
\mathcal{B}^{\scriptscriptstyle(z)}:=\big\{\,|z,j\rangle
\,|\,j\in\mathbb{Z}_d\big\}\,,
\ \mbox{where}\ \
XZ^z|z,j\rangle=\omega^j\,|z,j\rangle\,.
\end{equation}
Eigenvalues of every non-identity ${X^xZ^z}$ are distinct powers of $\omega$ \cite{Englert01,Schwinger60}.

With an integral power [obtained by repeatedly using~\eqref{Weyl commutation}] 
\begin{equation}
\label{(X^xZ^z)^k}
(XZ^z)^k=
\omega^{\frac{k(k-1)}{2}z} X^{k}Z^{kz}
\end{equation}
and the eigenvalue equation in~\eqref{B^z},
one can arrive at the spectral decomposition
\begin{equation}
\label{X^{k}Z^{kz}}
X^{k}Z^{kz}=\omega^{-\frac{k(k-1)}{2}z}
\sum_{j\,\in\,\mathbb{Z}_d}\,
\omega^{kj}\,
|z,j\rangle\langle z,j|
\end{equation}
of every operator in the subset $\mathcal{C}^{(1,z)}$. 
Now taking \eqref{X^{k}Z^{kz}} and \eqref{Z_i}, we can pronounce the average
values as
\begin{eqnarray}
\label{<X^{k}Z^{kz}>}
\langle X^{k}Z^{kz}\rangle_\rho&=&
\omega^{-\frac{k(k-1)}{2}z}
\sum_{j\,\in\,\mathbb{Z}_d}\,
\omega^{kj}\,
p^{\scriptscriptstyle(z)}_j
\quad\mbox{and}\\
\label{<Z^{k}>}
\langle Z^k\rangle_\rho&=&
\sum_{j\,\in\,\mathbb{Z}_d}\,
\omega^{kj}\,p_j\,,
\quad\mbox{where}\\
\label{pj}
p^{\scriptscriptstyle(z)}_j&=&\langle z, j|\,\rho\,|z,j\rangle
\quad\mbox{and}\quad
p_j=\langle j|\,\rho\,|j\rangle\qquad
\end{eqnarray}
are the probabilities for projective measurements in ${d+1}$ MUBs~\eqref{d+1 bases}.
Next, we can rewrite \eqref{tr_rho2-exp} as
\begin{eqnarray}
\label{tr_rho2-exp2}
\text{tr}(\rho^2)
&=&\tfrac{1}{d}\Big[1+\sum\limits_{z\,\in\,\mathbb{Z}_d}
\underbrace{\sum\limits_{k=1}^{d-1}
	{|\langle X^{k}Z^{kz} \rangle|}^{\,2}}%
_{\textstyle d \sum_{j}\big(p^{\scriptscriptstyle(z)}_j\big)^2-1}+
\underbrace{\sum\limits_{k=1}^{d-1}
	{|\langle Z^k \rangle|}^{2}}%
_{\textstyle d\sum_{j}(p_j)^2-1}
\Big]\quad
\nonumber\\
&=&
\label{tr_rho2-p}
\sum_{z\,\in\,\mathbb{Z}_d}\,
\sum_{j\,\in\,\mathbb{Z}_d}
\big(p^{\scriptscriptstyle(z)}_j\big)^2
+
\sum_{j\,\in\,\mathbb{Z}_d}
(p_j)^2-1\,.
\end{eqnarray}
Expression \eqref{tr_rho2-p} 
is achieved with the help of \eqref{<X^{k}Z^{kz}>}--\eqref{pj},
\begin{equation}
\label{sum pj=1}
\sum_{j\,\in\,\mathbb{Z}_d}
p^{\scriptscriptstyle(z)}_j=1=
\sum_{j\,\in\,\mathbb{Z}_d}
p_j
\end{equation}
[due to \eqref{p-const1}] for every $z$, and
${\textstyle\sum\nolimits_{k=0}^{d-1} \omega^{k(j-j')}=d\,\delta_{j,j'}}$.

Owing to ${\text{tr}(\rho^2)\leq1}$ [see \eqref{0<=S_2}], we reach the \emph{quadratic} QC for the Weyl operators in~\eqref{tr_rho2-exp} and thus 
\begin{equation}
\label{p^2<=1}
\sum_{z\,\in\,\mathbb{Z}_d}\,
\sum_{j\,\in\,\mathbb{Z}_d}
\big(p^{\scriptscriptstyle(z)}_j\big)^2
+
\sum_{j\,\in\,\mathbb{Z}_d}
(p_j)^2\,\leq2
\end{equation}
for the MUB-probabilities.
In \cite{Larsen90,Ivanovic92}, inequality~\eqref{p^2<=1} is achieved from ${\text{tr}(\rho^2)\leq1}$ via a different method (see also \cite{Klappenecker05}).
Using their result, that is \eqref{p^2<=1}, two tight URs are obtained in
\cite{Sanchez-Ruiz95,Ballester07} for ${d+1}$ MUBs.
In the case of ${d=2}$, these relations become \eqref{H-UR-JxJyJz} and \eqref{H2-UR-JxJyJz}.
For the cubic QC due to~\eqref{0<=S_3}, we need to express \eqref{tr_rho3-exp}
in terms of the probabilities.
In the next section, \eqref{tr_rho3-exp} is explicitly given for a qutrit.

Higher degree QCs for the Weyl operators and for the MUBs can be achieved---from~\eqref{0<=S_n}---by adopting the general formalism of Sec.~\ref{sec:QC}
like above.
The Weyl group exists for every $d$ \cite{Weyl32,Durt10,Englert06}, whereas a maximal set of ${d+1}$ MUBs is only known for a prime power dimension \cite{Wootters89,Bandyopadhyay02,Durt10}.
MUBs are \emph{optimal} for the quantum state estimation \cite{Ivanovic81,Wootters89}, where the QCs can be employed for
the validation of an estimated state.

\section{Qutrit and spin-1 system}\label{sec:qutrit}

In the case of ${d\geq3}$, there is a
\emph{cubic} QC as a result of \eqref{0<=S_3}.
For a qutrit (${d=3}$), let us first express $\text{tr}(\rho^m)$ of \eqref{tr(rho^m)-stnd} for ${m=1,2,3}$:
\begin{eqnarray}
\label{tr(rho)}
\text{tr}(\rho)&=&
\texttt{r}_{00}+\texttt{r}_{11}+\texttt{r}_{22}\,,\\
\label{tr(rho^2)}
\text{tr}(\rho^2)&=&
{\texttt{r}_{00}}^2+{\texttt{r}_{11}}^2+{\texttt{r}_{22}}^2+\nonumber\\
&& 2\,\big(\,|\texttt{r}_{01}|^2+|\texttt{r}_{02}|^2+|\texttt{r}_{12}|^2\,\big)\,,\quad\mbox{and}\quad
\\
\label{tr(rho^3)}
\text{tr}(\rho^3)&=&
{\texttt{r}_{00}}^3+{\texttt{r}_{11}}^3+{\texttt{r}_{22}}^3+\nonumber\\
&& 3\,\texttt{r}_{00}\,\big(\,|\texttt{r}_{01}|^2+|\texttt{r}_{02}|^2\,\big)+\nonumber\\
&& 3\,\texttt{r}_{11}\,\big(\,|\texttt{r}_{01}|^2+|\texttt{r}_{12}|^2\,\big)+\nonumber\\
&& 3\,\texttt{r}_{22}\,\big(\,|\texttt{r}_{02}|^2+|\texttt{r}_{12}|^2\,\big)+\nonumber\\
&& 3\,\big(\,\texttt{r}_{01}\,\texttt{r}_{12}\,\texttt{r}_{20}+ \overline{\texttt{r}}_{01}\,\overline{\texttt{r}}_{12}\,\overline{\texttt{r}}_{20}\,\big)
\end{eqnarray}
[for $\texttt{r}_{jk}$, see \eqref{r_kj}].
Here we consider two sets of operators: set~\eqref{Unitary-Basis} of the Weyl operators for a qutrit and a set of spin-1 operators.
In the following, we demonstrate:
how to achieve $\text{tr}(\rho^m)$, straight from~\eqref{tr(rho)}--\eqref{tr(rho^3)},
in terms of the expectation values of operators
in a given set without exploiting their algebraic properties.
Then, one gains automatically all the QCs  from~\eqref{1=S_1}--\eqref{0<=S_3}.

In \eqref{X_i} and \eqref{Z_i}, the Weyl operators are expressed in the linear combinations of operators belong to standard basis~\eqref{Stnd-Basis}.
Now we write
\begin{eqnarray}
\label{proj-op}
|j\rangle\langle k|
&=&X^j\,|0\rangle\langle 0|\,X^{-k}
=X^j\left[\tfrac{1}{d}
\sum_{z\,\in\,\mathbb{Z}_d} Z^z\,
\right ] X^{-k}\nonumber\\
&=& \tfrac{1}{d}
\sum_{z\,\in\,\mathbb{Z}_d}\, \omega^{-kz}\, X^{j-k}\,Z^z
\end{eqnarray}
by using \eqref{X_i}, \eqref{Z_i}, and \eqref{Weyl commutation}; see 
also \cite{Durt10}.
According to Born's rule~\eqref{Born rule-1}, the mean value is a linear function of an operator, so we own every $\texttt{r}_{kj}$ of \eqref{r_kj} as a linear sum of ${\langle X^xZ^z\rangle_\rho}$ through \eqref{proj-op}.
This constitutes a matrix equation such as \eqref{expt-values-std}.
By substituting $\texttt{r}_{kj}$ with
the associated linear combination in \eqref{tr(rho)}--\eqref{tr(rho^3)}, one can achieve~$\text{tr}(\rho^m)$ 
in terms of ${\langle X^x\,Z^z\rangle}$ for a qutrit:
\begin{eqnarray}
\label{tr_rho3-exp-qutrit}
&&\text{tr}(\rho^3)=\nonumber\\
&&\tfrac{1}{9}\,\big[\,
1+\langle X\rangle^3+ \langle X^2\rangle^3+\langle XZ\rangle^3+
\langle X^2Z^2\,\rangle^3+\nonumber\\
&&
\qquad\quad
\langle XZ^2\rangle^3+ \langle X^2Z\rangle^3+\langle Z\rangle^3+
\langle Z^2\,\rangle^3+\nonumber\\
&&\qquad
6\,\big(\,|\langle X\rangle|^2+|\langle XZ\rangle|^2
+|\langle XZ^2\rangle|^2+|\langle Z\rangle|^2\,\big)
\nonumber\\
&&
-\,3\,\big(
\langle X\rangle\langle XZ\rangle\langle XZ^2\rangle+
\langle X^2\rangle\langle X^2Z\rangle\langle X^2Z^2\rangle+\nonumber\\
&&\qquad
\langle Z\rangle\langle XZ\rangle\langle X^2Z\rangle+
\langle Z^2\rangle\langle XZ^2\rangle\langle X^2Z^2\rangle+\nonumber\\
&&\qquad
\omega\,\langle Z\rangle\langle X^2\rangle\langle XZ^2\rangle+
\omega\,\langle Z^2\rangle\langle X\rangle\langle X^2Z\rangle+\nonumber\\
&&\qquad
\omega^2\langle Z\rangle\langle X\rangle\langle X^2Z^2\rangle+
\omega^2\langle Z^2\rangle\langle X^2\rangle\langle XZ\rangle
\big)
\big]\,,\qquad\quad
\end{eqnarray}
where $\omega=\exp(\text{i}\tfrac{2\pi}{3})$, and the term $6(\cdots)$ is
$3(3\text{tr}(\rho^2)-1)$.
In Sec.~\ref{sec:unitary-basis}, we get \eqref{tr_rho2-exp} and \eqref{tr_rho3-exp} from \eqref{tr(rho^m)}
by exploiting algebraic properties~\eqref{Weyl commutation} and \eqref{traceless}.
One can compare that both the methods deliver the same items.

The next example,
a spin-1 particle is a ${d=3}$ levels quantum system (qutrit) if we consider only the spin degree of freedom.
Here we take a set of three Hermitian operators from Chap.~7 in \cite{Peres95}:
\begin{eqnarray}
	\label{Jx}
J_x&:=&-\text{i}\big(|0\rangle\langle 1|-|1\rangle\langle 0|\big)\,,
	\\
	\label{Jy}
J_y&:=&-\text{i}\big(|0\rangle\langle 2|-|2\rangle\langle 0|\big)\,,
	\quad\mbox{and}
	\\
	\label{Jz}
J_z&:=&-\text{i}\big(|1\rangle\langle 2|-|2\rangle\langle 1|\big)\,.	
\end{eqnarray}
They obey the commutation relation ${J_xJ_y-J_yJ_x=\text{i}J_z}$ plus those obtained by the cyclic permutations of ${x,y,z}$, and thus they represent ${\text{spin-}1}$ observables. 
One can check that 
${J_x, J_y,J_z}$ with ${J_x^{\,2},J_y^{\,2},J_z^{\,2}}$ and the 
anticommutators
\begin{equation}
	\label{anti-commutators}
	K_{xy}=J_xJ_y+J_yJ_x\,,\quad
	K_{yz}\quad\mbox{and}\quad\,K_{zx}
\end{equation}
(attain by the cyclic permutations)
constitute a set of nine linearly independent operators, hence they form a Hermitian-basis of $\mathscr{B}(\mathscr{H}_3)$.
Though it is not an orthonormal basis with respect to inner product~\eqref{HS-inner-pro}.

One can recognize that ${J_x,J_y,J_z}$ and ${K_{xy},K_{yz},K_{zx}}$
are the Gell-Mann operators \cite{Gell-Mann61},
but ${J_x^{\,2},J_y^{\,2},J_z^{\,2}}$ are not.
We want to emphasize that the QCs on their average values 
can be derived from \cite{Kimura03,Byrd03}.
So the following analysis is merely an
alternative procedure that does not require the Lie algebra of $SU(3)$.

After expressing the elements of standard basis~\eqref{Stnd-Basis} 
in terms of the spin operators, we can write the average values as
\begin{eqnarray}
\label{<J>}
\texttt{r}_{00}&=&\tfrac{1}{2}\left(\hphantom{-}\langle J_x^{\,2}\rangle+
\langle J_y^{\,2}\rangle-\langle J_z^{\,2}\rangle\right)\,,
\nonumber\\
\texttt{r}_{11}&=&\tfrac{1}{2}\left(\hphantom{-}\langle J_x^{\,2}\rangle-
\langle J_y^{\,2}\rangle+\langle J_z^{\,2}\rangle\right)\,,
\nonumber\\
\texttt{r}_{22}&=&\tfrac{1}{2}\left(-\langle J_x^{\,2}\rangle+
\langle J_y^{\,2}\rangle+\langle J_z^{\,2}\rangle\right)\,,
\\
\texttt{r}_{01}&=&\tfrac{1}{2}\left(\hphantom{-}\langle K_{yz}\rangle-
\text{i}\,\langle J_x\rangle\right)=\overline{\texttt{r}}_{10}\,,
\nonumber\\
\texttt{r}_{02}&=&\tfrac{1}{2}\left(-\langle K_{zx}\rangle-
\text{i}\,\langle J_y\rangle\right)=\overline{\texttt{r}}_{20}\,,
\quad\mbox{and}
\nonumber\\
\texttt{r}_{12}&=&\tfrac{1}{2}\left(\hphantom{-}\langle K_{xy}\rangle-
\text{i}\,\langle J_z\rangle\right)=\overline{\texttt{r}}_{21}\,.
\nonumber
\end{eqnarray}
This set of equations frames a 
matrix equation of the kind in \eqref{expt-values-std}.
Employing Eqs.~\eqref{<J>}, we can rephrase \eqref{tr(rho)}--\eqref{tr(rho^3)}
as
\begin{eqnarray}
\label{tr(rho)-J}
\text{tr}(\rho)&=&\tfrac{1}{2}\left(\langle J_x^{\,2}\rangle+
\langle J_y^{\,2}\rangle+\langle J_z^{\,2}\rangle\right)\,,
\\
\label{tr(rho^2)-J}
\text{tr}(\rho^2)&=&-1+
\langle J_x^{\,2}\rangle^2+
\langle J_y^{\,2}\rangle^2+\langle J_z^{\,2}\rangle^2+\nonumber\\
&&\quad
\tfrac{1}{2}\big(\langle J_x\rangle^2+\langle J_y\rangle^2+
\langle J_z\rangle^2+\nonumber\\
&&\qquad
\langle K_{xy}\rangle^2+\langle K_{yz}\rangle^2+\langle K_{zx}\rangle^2
\big)\,,\ \mbox{and}\qquad\
\\
\label{tr(rho^3)-J}
\text{tr}(\rho^3)&=&1-
3\,\langle J_x^{\,2}\rangle\langle J_y^{\,2}\rangle\langle J_z^{\,2}\rangle+
\nonumber\\
&&\
\tfrac{3}{4}\,\big[\,
\left(\langle J_x\rangle^2+\langle K_{yz}\rangle^2\right)\langle J_x^{\,2}\rangle+
\nonumber\\
&&\quad\ \
\left(\langle J_y\rangle^2+\langle K_{zx}\rangle^2\right)\langle J_y^{\,2}\rangle+
\nonumber\\
&&\quad\ \
\left(\langle J_z\rangle^2+\langle K_{xy}\rangle^2\right)\langle J_z^{\,2}\rangle
\nonumber\\
&&\
-\,\langle K_{xy}\rangle\langle K_{yz}\rangle\langle K_{zx}\rangle+
\langle J_{x}\rangle\langle K_{xy}\rangle\langle J_{y}\rangle
\nonumber\\
&&\quad
+\,\langle J_{y}\rangle\langle K_{yz}\rangle\langle J_{z}\rangle+
\langle J_{z}\rangle\langle K_{zx}\rangle\langle J_{x}\rangle
\,\big]\,.
\end{eqnarray}
Here, in each example, one can clearly perceive $\text{tr}(\rho^2)$ and $\text{tr}(\rho^3)$ as quadratic and cubic polynomials of the mean values.
Plugging \eqref{tr(rho)-J}--\eqref{tr(rho^3)-J} in \eqref{1=S_1}--\eqref{0<=S_3}, one captures all the---linear, quadratic, and cubic---QCs for the spin-1 operators.
The linear constraint ${\langle J_x^{\,2}+J_y^{\,2}+J_z^{\,2}\rangle=\langle 2I\rangle}$
is used to get \eqref{tr(rho^2)-J} and \eqref{tr(rho^3)-J}
in the above forms.

Now, let us call 
$J_x,J_y,J_z,K_{yz},K_{zx},K_{xy},
J_x^{\,2},J_y^{\,2},J_z^{\,2}$
as ${A_1,\cdots,A_9}$, respectively.
In this case, every pure state $\rho_\text{pure}=|\psi\rangle\langle\psi|$ [for ${|\psi\rangle}$, see \eqref{|psi>}] of a qutrit delivers an extreme point of
the allowed region $\mathcal{E}$, and the extreme points can be parameterized as
\begin{eqnarray}
\label{J-pure}
\langle A_1\rangle_{\rho_\text{pure}}&=&
\sin 2\theta_0 \cos\theta_1 \sin\phi_1\,,
\nonumber\\
\langle A_2\rangle_{\rho_\text{pure}}&=&
\sin 2\theta_0 \sin\theta_1 \sin\phi_2\,,
\nonumber\\
\langle A_3\rangle_{\rho_\text{pure}}&=&
-(\sin\theta_0)^2 \sin 2\theta_1 \sin(\phi_1-\phi_2)\,,
\nonumber\\
\langle A_4\rangle_{\rho_\text{pure}}&=&
\sin 2\theta_0 \cos\theta_1 \cos\phi_1\,,
\nonumber\\
\langle A_5\rangle_{\rho_\text{pure}}&=&
-\sin 2\theta_0 \sin\theta_1 \cos\phi_2\,,
\\
\langle A_6\rangle_{\rho_\text{pure}}&=&
(\sin\theta_0)^2 \sin 2\theta_1 \cos(\phi_1-\phi_2)\,,
\nonumber\\
\langle A_7\rangle_{\rho_\text{pure}}&=&
(\cos\theta_0)^2+(\sin\theta_0)^2(\cos\theta_1)^2\,,
\nonumber\\
\langle A_8\rangle_{\rho_\text{pure}}&=&
(\cos\theta_0)^2+(\sin\theta_0)^2(\sin\theta_1)^2\,,
\quad\mbox{and}\qquad
\nonumber\\
\langle A_9\rangle_{\rho_\text{pure}}&=&
(\sin\theta_0)^2\,,
\nonumber
\end{eqnarray}
where ${\theta_0,\theta_1\in[0,\tfrac{\pi}{2}]}$ and ${\phi_1,\phi_2\in[0,2\pi)}$.
By putting expectation values \eqref{J-pure} in \eqref{tr(rho)-J}--\eqref{tr(rho^3)-J}, one can verify that ${\text{tr}(\rho_\text{pure}^{\,m})=1}$ for all $m=1,2,$ and 3.

The minimum and maximum eigenvalues of everyone in ${\{A_1,\cdots,A_6\}}$ are $-1$ and $+1$ and of each one in
${\{A_7,A_8,A_9\}}$ are 0 and $1$, respectively. Taking \eqref{Adot}--\eqref{umax(A)}, we formulate uncertainty or certainty measures for $\{A_i\}_{i=1}^9$, and a few combined measures are listed in
\begin{eqnarray}
\label{H-J}
6\ln 2 &\leq& \sum_{i=1}^{9}H(\langle A_i\rangle)\,,
\\
\label{u1/2-J}
3 + 6 \sqrt{2} &\leq& \sum_{i=1}^{9}u_{\sfrac{1}{2}}(\langle A_i\rangle)\,,
\\
\label{u2-J}
&& \sum_{i=1}^{9}u_{2}(\langle A_i\rangle)\leq 6\,,\quad\mbox{and}
\\
\label{umax-J}
&& \sum_{i=1}^{9}u_{\text{max}}(\langle A_i\rangle)\leq 6.51702\,.
\end{eqnarray}
As described in Sec.~\ref{sec:QC}, we find the absolute minimum of a concave function and 
maximum of a convex function by putting \eqref{J-pure} in the above functions
and changing the four parameters $\theta$'s and $\phi$'s.
As a result, we achieve tight URs \eqref{H-J} and \eqref{u1/2-J} and CRs \eqref{u2-J} and \eqref{umax-J} for the nine spin-1 observables.
The basis ${\mathcal{B}=\{|0\rangle,|1\rangle,|2\rangle\}}$ in \eqref{B_i} is a common eigenbasis of ${\{A_7,A_8,A_9\}}$, a qutrit's state 
$\rho=|j\rangle\langle j|$ that corresponds to a ket in $\mathcal{B}$ saturates inequalities \eqref{H-J}--\eqref{u2-J}.
One pure state that saturates CR \eqref{umax-J}, the corresponding parameters are
\begin{eqnarray}
\label{umax-ang}
\theta_0 = 0.482720\,,&&\quad
\theta_1 = 0.785398\,,\nonumber\\
\phi_1 = 2.520428\,,&&\quad
\phi_2 = 3.762757\,.
\end{eqnarray}

Since the square of every operator in the set ${\{A_i\}_{i=1}^9}$ lies in the set,
\begin{eqnarray}
\label{A^2}
(A_1)^2&=&(A_4)^2=(A_7)^2=A_7\,,\nonumber\\
(A_2)^2&=&(A_5)^2=(A_8)^2=A_8\,,\quad\mbox{and}\\
(A_3)^2&=&(A_6)^2=(A_9)^2=A_9\,,\nonumber
\end{eqnarray}
a sum of (the square of) the standard deviations 
${\Delta A_i}$ [see \eqref{std-dev}] acts
as a concave function on the allowed region for the set.
As above we reach the global minima and thus establish the tight URs
\begin{eqnarray}
\label{std-J 9}
4 &\leq& \sum_{i=1}^{9}\Delta A_i\,,\\
\label{std-J 6}
1 + 2\sqrt{2} &\leq& \sum_{i=1}^{6}\Delta A_i\,,
\\
\label{aq-std-J}
\tfrac{10}{3} &\leq& \sum_{i=1}^{9}\big(\Delta A_i\big)^2\,,
\quad\mbox{and}\quad
\tfrac{8}{3} \leq \sum_{i=1}^{6}\big(\Delta A_i\big)^2\,.
\qquad\quad
\end{eqnarray}
URs \eqref{std-J 9} and \eqref{std-J 6} are saturated by
the eigenstates of $A_i$, $i=1,\cdots,6$, 
associated with 0 and the non-zero eigenvalues, respectively.
The null-space (eigenspace associated with 0) of $A_i$
is the linear span of a ket in $\mathcal{B}$.
The equal superposition kets 
${\tfrac{1}{\sqrt{3}}(|0\rangle+e^{\text{i}\phi_1}|1\rangle+e^{\text{i}\phi_2}|2\rangle)}$ provide the minimum uncertainty (pure) states for both the URs in \eqref{aq-std-J}.


\section{Spin-$\mathsf{j}$ operators}\label{sec:spin-j}

A spin-\textsf{j} particle is a quantum system of ${d=2\,\mathsf{j}+1}$ levels provided we consider only the spin degree of freedom, and \textsf{j} can be 
${\tfrac{1}{2}, 1,\tfrac{3}{2},2,\cdots\,}$.
Let us take the spin-\textsf{j} operators
${J_x=\tfrac{1}{2}(J_++J_-)}$, 
${J_y=\tfrac{1}{2\text{i}}(J_+-J_-)}$, and $J_z$
whose actions on the eigenbasis 
${\{|\mathsf{m}\rangle : \mathsf{m=j,j}-1,\cdots,-\mathsf{j}\}}$ of $J_z$
are described as
\begin{eqnarray}
\label{JpmJz}
J_\pm\,|\mathsf{m}\rangle&=&
\sqrt{\mathsf{(j\mp m)(j\pm m}+1)}\;
|\mathsf{m}\pm 1\rangle\quad\mbox{and}\quad\\
J_z\,|\mathsf{m}\rangle&=&\mathsf{m}\,|\mathsf{m}\rangle\,.
\end{eqnarray}
For ${\mathsf{j}=\tfrac{1}{2}}$, the vector operator ${\vec{J}:=(J_x,J_y,J_z)}$	
is the same as	the Pauli vector operator ${\vec{\sigma}:=(X,Y,Z)}$ in Appendix
up to a factor $\tfrac{1}{2}$.
In \eqref{Jx}--\eqref{Jz}, the spin-1 operators are represented
in the common eigenbasis $\mathcal{B}$ of 
${\{J_x^{\,2},J_y^{\,2},J_z^{\,2}\}}$.

The permitted region $\mathcal{E}$ for the three spin-observables is bounded by the QC
\begin{equation}
\label{JxJyJz,QC}
{\langle J_x\rangle}^2+{\langle J_y\rangle}^2+{\langle J_z\rangle}^2\leq \mathsf{j}^{\,2}\,,
\end{equation}
which says that the length of the vector 
${(\langle J_x\rangle,\langle J_y\rangle,\langle J_z\rangle)}$
cannot be more than $\mathsf{j}$ \cite{Atkins71}.
So $\mathcal{E}$ is the closed ball of radius $\mathsf{j}$ in hyperrectangle~\eqref{hyperrectangle} that is the cube $[-\mathsf{j},\mathsf{j}]^{\times 3}$ here.
Note that, except ${\mathsf{j}=\tfrac{1}{2}}$, an interior point of $\mathcal{E}$ corresponds to \emph{not one but many} (pure as well as mixed) quantum states.
However, every extreme point of $\mathcal{E}$ comes from a unique pure state 
${\chi(\alpha,\beta)=|\alpha,\beta\rangle\langle\alpha,\beta|}$, where
\begin{equation}
\label{bloch-ket}
|\alpha,\beta\rangle=\sum_{\mathsf{m=-j}}^{\mathsf{j}}
{\scriptstyle\sqrt{\tfrac{(2\mathsf{j})!}{(\mathsf{j+m})!\,(\mathsf{j-m})!}}
\left(\cos\tfrac{\alpha}{2}\right)^{\mathsf{j+m}}
\left(\sin\tfrac{\alpha}{2}\right)^{\mathsf{j-m}}
e^{-\text{i}\mathsf{m}\beta}}|\mathsf{m}\rangle
\end{equation}
is known as the angular momentum (or atomic) coherent state-vector \cite{Atkins71,Arecchi72}.
With ${J_x^{\,2}+J_y^{\,2}+J_z^{\,2}=\mathsf{j(j}+1)I}$, QC~\eqref{JxJyJz,QC} can be turned into a tight UR
\begin{equation}
\label{JxJyJz,UR}
\mathsf{j}\leq 
(\Delta J_x)^2+(\Delta J_y)^2+(\Delta J_z)^2\,,
\end{equation}
for which all the coherent states are the minimum uncertainty states (see Chap.~10 in \cite{Peres95}).
UR~\eqref{JxJyJz,UR} is also captured in \cite{Larsen90,Abbott16,Hofmann03,Dammeier15}.
In fact, \eqref{JxJyJz,QC} can also be interpreted as CR because on the left-hand-side 
there is a convex function of the expectation values.
In \cite{Dammeier15},
$(\Delta\, \widehat{\eta}.\vec{J}\,)^2$ is studied as a function of the unit vector ${\widehat{\eta}\in\mathbb{R}^3}$ for a fixed state $\rho$, and then the uncertainty regions of
${((\Delta J_x)^2,(\Delta J_y)^2,(\Delta J_z)^2)}$ are plotted by taking all $\rho$'s. Various URs are also obtained there for the three operators ${J_x,J_y,}$ and $J_z$. 
Our regions $\mathcal{E}$ and $\mathcal{R}$'s are different from the uncertainty regions: $\mathcal{E}$ and $\mathcal{R}$
are in the space of expectation values, and both are convex sets.

We can parametrize the extreme points of $\mathcal{E}$ as
\begin{eqnarray}
\label{JxJyJz-para}
\langle\alpha,\beta|J_x|\alpha,\beta\rangle&=&\mathsf{j}\,\sin\alpha\cos\beta\,,
\nonumber\\
\langle\alpha,\beta|J_y|\alpha,\beta\rangle&=&\mathsf{j}\,\sin\alpha\sin\beta\,,
\\
\langle\alpha,\beta|J_z|\alpha,\beta\rangle&=&\mathsf{j}\,\cos\alpha\,,
\nonumber
\end{eqnarray}
where ${\alpha\in[0,\pi]}$ and ${\beta\in[0,2\pi)}$, and can define different uncertainty or certainty measures  on $\mathcal{E}$ using \eqref{Adot}--\eqref{umax(A)}.
Since the minimum and maximum eigenvalues of $J_i$ for every ${i=x,y,z}$ are $-\mathsf{j}$ and $+\mathsf{j}$, respectively,
\begin{equation}
\label{Jdot}
\langle\dot{J}_i\,\rangle=
\tfrac{1}{2}\big(1-\tfrac{\langle J_i\rangle}{\mathsf{j}}\big)
\quad\mbox{and}\quad
\langle\mathring{J}_i\,\rangle=
\tfrac{1}{2}\big(1+\tfrac{\langle J_i\rangle}{\mathsf{j}}\big)\,,
\end{equation}	
which
are functions of $\alpha$ and $\beta$ on the sphere specified by \eqref{JxJyJz-para}.
By varying the two angles we reach the tight lower and upper bounds of the uncertainty and certainty measures presented as follows
\begin{eqnarray}
\label{H-UR-JxJyJz}
2\ln 2&\leq&\sum_{i=x,y,z}H(\langle J_i\rangle)\,,\\
\label{H2-UR-JxJyJz}
3\ln(\tfrac{3}{2})&\leq&\sum_{i=x,y,z}H_2(\langle J_i\rangle)\,,\\
\label{u-UR-JxJyJz}
1+2\sqrt{2}&\leq& \sum_{i=x,y,z} 
u_{\sfrac{1}{2}}(\langle J_i\rangle)\,,\\
\label{u2-UR-JxJyJz}
&& \sum_{i=x,y,z}u_2(\langle J_i\rangle)\leq 2 \,,
\quad \mbox{and}\qquad\\
\label{umax-UR-JxJyJz}
&& \sum_{i=x,y,z}u_\text{max}(\langle J_i\rangle)\leq 
\tfrac{1}{2}(3+\sqrt{3}) \,,
\end{eqnarray}
where ${H_2=-\ln(u_2)}$ is like the R\'{e}nyi entropy \cite{Renyi61}
of order~2.

\begin{figure}
	\centering
	\includegraphics[width=0.45\textwidth]{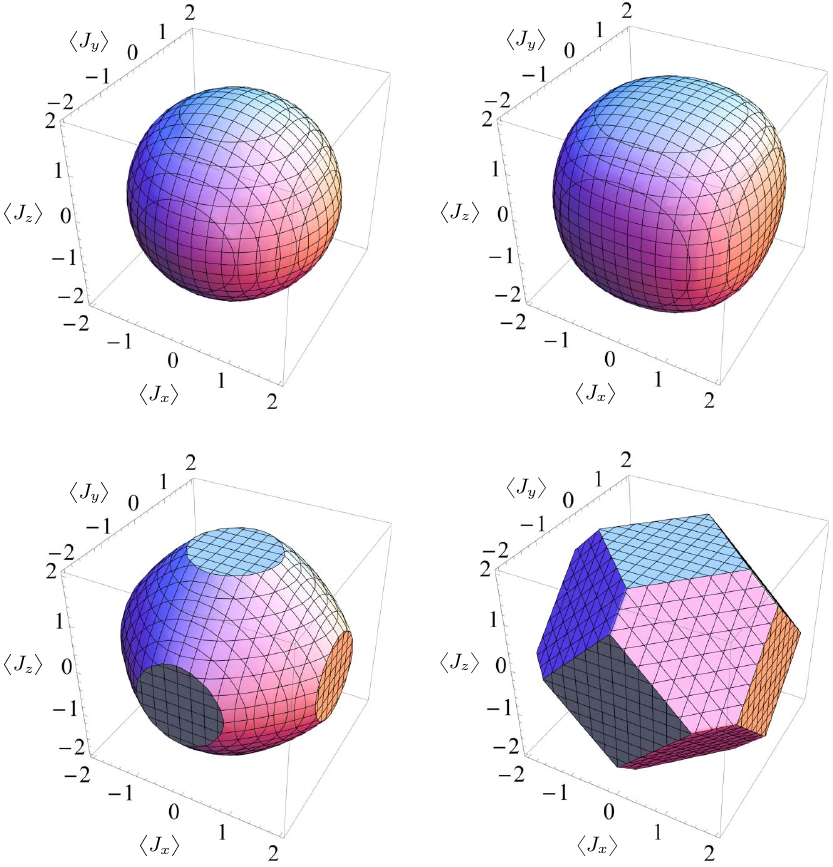}
	\caption{From top-left to bottom-right, along the rows, the first region is the permissible region $\mathcal{E}$ bounded by QC \eqref{JxJyJz,QC} and
	the second one is $\mathcal{R}^{\text{spin}}_H$. The third and fourth regions are $\mathcal{R}^{\text{spin}}_{H_2}$ and $\mathcal{R}^{\text{spin}}_{u_{\text{max}}}$, respectively.
	Although these regions are plotted for ${\mathsf{j}=2}$,
	they will be of the same shapes in the cube $[-\mathsf{j},\mathsf{j}]^{\times 3}$ for other $\mathsf{j}$-values.}
	\label{fig:regions-MUBs} 
\end{figure}

All \eqref{H-UR-JxJyJz}--\eqref{umax-UR-JxJyJz}
hold for every ${\mathsf{j}=\tfrac{1}{2}, 1,\tfrac{3}{2},2,\cdots\,}$
and hence in every dimension ${d=2\,\mathsf{j}+1}$, and they are saturated by some
angular momentum coherent states $\chi(\alpha,\beta)$.
Like \eqref{R_H}, the regions characterized by URs \eqref{H-UR-JxJyJz}, \eqref{H2-UR-JxJyJz}, and CR \eqref{umax-UR-JxJyJz} are denoted here by $\mathcal{R}^{\text{spin}}_H$, $\mathcal{R}^{\text{spin}}_{H_2}$, and $\mathcal{R}^{\text{spin}}_{u_{\text{max}}}$, respectively.
Along with $\mathcal{E}$, they are
displayed in Fig.~\ref{fig:regions-MUBs} for ${\mathsf{j}=2}$.
$\mathcal{E}$ resides in every $\mathcal{R}$, and one can also perceive that ${\mathcal{R}^{\text{spin}}_{H_2}\subset\mathcal{R}^{\text{spin}}_{u_{\text{max}}}}$.
We can not right away say which of the tight URs, \eqref{H-UR-JxJyJz} or \eqref{H2-UR-JxJyJz}, is superior because neither
$\mathcal{R}^{\text{spin}}_H$ is completely contained in $\mathcal{R}^{\text{spin}}_{H_2}$ nor vice versa.
Similarly, it is difficult to compare \eqref{H-UR-JxJyJz} and \eqref{umax-UR-JxJyJz}
as ${\mathcal{R}^{\text{spin}}_H\nsubset\mathcal{R}^{\text{spin}}_{u_{\text{max}}}}$
and
${\mathcal{R}^{\text{spin}}_H\nsupset\mathcal{R}^{\text{spin}}_{u_{\text{max}}}}$.
If one region is not a subset of other then one can take the area of a region as a figure of merit to compare different CRs and/or URs.
However, in the paper, mostly those cases are reported where one region is completely submerged in another.

Since \eqref{JxJyJz,QC} and \eqref{u2-UR-JxJyJz} are the same, every 
angular momentum coherent state saturates \eqref{u2-UR-JxJyJz}.
$\mathcal{E}$ touches the periphery of $\mathcal{R}^{\text{spin}}_H$
at six different points that are related to eigenstates of $J_x,J_y,J_z$ corresponding to their extreme-eigenvalues $\pm\,\mathsf{j}$.
These six pure states are only the minimum uncertainty states for UR~\eqref{H-UR-JxJyJz} as well as
UR~\eqref{u-UR-JxJyJz}.  
The eight coherent states ${\chi(\alpha,\beta)}$---for which $\alpha=\arccos(\tfrac{1}{\sqrt{3}})$ and
${\beta=\tfrac{\pi}{4},\tfrac{3\pi}{4},\tfrac{5\pi}{4},\tfrac{7\pi}{4}}$, and the remaining four can be obtained by changing $\alpha$ into ${\pi-\alpha}$ and
$\beta$ into ${\pi+\beta\, (\text{mod}\, 2\pi)}$---saturate inequalities \eqref{H2-UR-JxJyJz} and \eqref{umax-UR-JxJyJz}.
The permitted region $\mathcal{E}$ touches the boundary of
$\mathcal{R}^{\text{spin}}_{H_2}$ and $\mathcal{R}^{\text{spin}}_{u_{\text{max}}}$
at the associated eight points.
The six cross sections in
$\mathcal{R}^{\text{spin}}_{H_2}$ and $\mathcal{R}^{\text{spin}}_{u_{\text{max}}}$ are due to ${-\mathsf{j}\leq\langle J_i\rangle\leq\mathsf{j}}$ required for every ${i=x,y,z}$.

In the case of ${\mathsf{j}=\tfrac{1}{2}}$, \eqref{JxJyJz,QC} and \eqref{|r|^2<=1}
are equal, $\mathcal{E}$ is the Bloch ball, and all the coherent states
become qubit's pure states.
Corresponding to the eight minimum uncertainty states for UR~\eqref{H2-UR-JxJyJz}, the Bloch vectors are
${\{\pm\widehat{a}_i\}_{i=1}^{4}}$ \cite{Wootters07,Appleby14},
where
\begin{equation}
\label{a1a2a3-v1v2v3}
\begin{pmatrix}
\widehat{a}_1\\
\widehat{a}_2\\
\widehat{a}_3
\end{pmatrix}
= \tfrac{1}{\sqrt{3}}
\begin{pmatrix}
\hphantom{-}1 & -1 & -1  \\
-1 & \hphantom{-}1 & -1  \\
-1 & -1 & \hphantom{-}1  
\end{pmatrix}
\begin{pmatrix}
\widehat{v}_1\ \\
\widehat{v}_2\ \\
\widehat{v}_3\
\end{pmatrix}
\end{equation}
and ${\widehat{a}_4=
	-\textstyle\sum\nolimits_{i=1}^{3}\widehat{a}_i}$
are given in the $v$-coordinate system [see Appendix].
One can easily deduce that both ${\{\widehat{\mathsf{a}}_i\}_{i=1}^{3}}$ presented in Appendix~\ref{subsec:3settings} and ${\{\widehat{a}_i\}_{i=1}^{3}}$
share the same Gram matrix, \eqref{SIC-POVM-Gram-matrix}.
The two sets of vectors are related by an invertible linear transformation that can be obtained by $\textbf{M}$ in 
\eqref{M^-1 SIC} and the square matrix in~\eqref{a1a2a3-v1v2v3}. 
Each of ${\{\widehat{\mathsf{a}}_i\}_{i=1}^{4}}$ and ${\{\widehat{a}_i\}_{i=1}^{4}}$ constitutes a SIC-POVM via \eqref{Pi} for a qubit \cite{Rehacek04},
and the later one is known as the Weyl-Heisenberg covariant SIC-POVM \cite{Appleby09}.

\section{Conclusion and outlook}\label{sec:conclusion}

There are three primary contributions from this article.
First, we provided a basis-independent systematic procedure to
obtain the QCs for any set of operators that act on a qudit's Hilbert space.
The QCs are necessary and sufficient restrictions 
that analytically specify the permitted region $\mathcal{E}$ of
the expectation values.
Second, we showed how to define uncertainty and certainty measures
on the allowed region $\mathcal{E}$, and their properties are discussed. 
With a straightforward mechanism---that is also employed in \cite{Riccardi17,Sehrawat17}---we
achieved tight CRs and URs.
Third, we bounded a regions $\mathcal{R}$ by a tight CR or UR in the space of expectation values and exhibited the gap $\mathcal{R}\setminus\mathcal{E}$ between 
$\mathcal{R}$ and the allowed region $\mathcal{E}$ through figures.
Our additional contributions are: 
(\textit{i}) the QCs for the Weyl operators and the spin observables are reported.
(\textit{ii}) Various tight URs and CRs are obtained for the spin-1 observables
as well as for ${\{J_x,J_y,J_z\}}$ in the case of an arbitrary spin 
${\mathsf{j}=\tfrac{1}{2}, 1,\tfrac{3}{2},2,\cdots\,}$.
Since all the extreme points of the permissible region
for ${\{J_x,J_y,J_z\}}$ come from the angular momentum coherent states,
always a coherent state is a minimum uncertainty state for the UR
formulated for the three observables.
(\textit{iii}) The case of a single qubit is thoroughly investigated in Appendix~\ref{sec:qubit} that includes Schr\"{o}dinger's UR, and tight URs CRs are presented there for the SIC-POVM.

Choice of an uncertainty measure to get a UR is a user's choice.
We have not yet found a single certainty or uncertainty measure 
that is better than others in the sense that it always provides 
a smaller region $\mathcal{R}$.
In some examples, one behaves better, whereas in another example
there is another.
To compare different CRs and/or URs, the area (or volume) of $\mathcal{R}$
can be a figure of merit, particularly when one region is not contained in another.
Although, it is not easy to compute such an area.

Naturally, $\mathcal{E}$ lies in all such $\mathcal{R}$'s, however it is not a primary objective of a UR to put a constraint on the mean values but on a combined uncertainty. 
To draw a comparison between the QCs and URs, first, we have to put them on an equal footing.
That may or may not be possible because
a QC is primarily a bound on expectation values not, generally, on a combined uncertainty.

URs play very important roles in different branches of physics and mathematics, recently they are applied in the field of quantum information
(see Sec.~VI in \cite{Coles17}).
One can employ the QCs for those purposes as well as for the quantum state estimation \cite{Paris04}, where one can directly appoint the QCs for the validation of an estimated state.

\begin{acknowledgments} 
	I am very grateful to Titas Chanda for crosschecking the numerical results.
\end{acknowledgments}

\appendix

\section{Qubit}\label{sec:qubit}

For a qubit (${d=2}$), 
the Pauli operators ${X,Y,Z}$ \cite{Pauli27} with 
the identity operator $I$
constitute the Hermitian-basis of $\mathscr{B}(\mathscr{H}_2)$
\cite{Kimura03,Byrd03}. 
The operators $X$ and $Z$ are defined in \eqref{X_i} and \eqref{Z_i}, respectively, and ${Y=\text{i}XZ}$.
In this basis, we can express qubit's state as 
\begin{equation}
\label{rho-in-IXYZ}
\rho=\tfrac{1}{2}\big(I+\langle X\rangle\, X +\langle Y\rangle\, Y
+\langle Z\rangle\, Z\big)\,,
\end{equation}
where ${\vec{r}:=(\langle X\rangle, \langle Y\rangle,\langle Z\rangle)\in\mathbb{R}^3}$ is the well-known Bloch vector \cite{Bloch46,Bengtsson06} that is the mean value of the Pauli vector operator ${\vec{\sigma}=(X,Y,Z)}$.
Conditions \eqref{1=S_1} and \eqref{0<=S_2} now become
$\langle I\rangle_\rho=1$ and 
\begin{equation}
\label{|r|^2<=1}
r^2:=|\vec{r}\,|^2=\langle X\rangle_\rho^2+\langle Y\rangle_\rho^2+\langle Z\rangle_\rho^2
\leq 1\,,
\end{equation}
respectively.

A projective measurement on a qubit can be completely specified by a three-component real unit vector \cite{A}.
So, we begin with three linearly independent unit vectors 
${\widehat{a},\widehat{b},\widehat{c}\in\mathbb{R}^3}$, and define three Hermitian operators
\begin{eqnarray}
\label{A-qubit}
A&\,:=\,&\widehat{a}\cdot\vec{\sigma}\,:=\,2\,|a\rangle\langle a|-I\,,
\\
\label{B-qubit}
B&\,:=\,&\widehat{b}\cdot\vec{\sigma}\,:=\,2\,|b\rangle\langle b|-I\,,
\quad\mbox{and}
\\
\label{C-qubit}
C&\,:=\,&\widehat{c}\cdot\vec{\sigma}\,:=\,2\,|c\rangle\langle c|-I\,.
\end{eqnarray}
One can check that ${A^2=I}$ [with \eqref{AB}], hence its eigenvalues are $\pm1$, and then ${\langle A\rangle_\rho\in[-1,1]}$ is due to \eqref{<A>in}.
By definition~\eqref{A-qubit}, ${|a\rangle}$ and 
${|a^\perp\rangle}$ (such that ${\langle a|a^\perp\rangle=0}$) are eigenkets of $A$ corresponding to the eigenvalues ${+1}$ and ${-1}$, respectively, and similarly for $B$ and $C$.

One can verify that the inner product between a pair of such operators is
\begin{equation}
\label{angle-bt-axis}
\lgroup A,B\,\rgroup_\textsc{hs}=
2\;\widehat{a}\cdot\widehat{b}=
4\;{|\langle a|b\rangle|}^2-2
\end{equation}
by using 
\begin{eqnarray}
\label{AB}
&&AB=(\widehat{a}\cdot\vec{\sigma})(\widehat{b}\cdot\vec{\sigma})
=(\widehat{a}\cdot\widehat{b})\,I+\text{i}(\widehat{a}\times\widehat{b})
\cdot\vec{\sigma}\,,\quad\\
\label{tr(pauli)}
&&\text{tr}(X)=\text{tr}(Y)=\text{tr}(Z)=0\,,
\quad\mbox{and}\quad
\text{tr}(I)=d=2\,,\qquad
\end{eqnarray}
where ${\widehat{a}\cdot\widehat{b}}$ and ${\widehat{a}\times\widehat{b}}$ are the dot and cross product.
Taking the statistical operator from~\eqref{rho-in-IXYZ} and applying \eqref{angle-bt-axis} and \eqref{tr(pauli)} to the Born rule, \eqref{Born rule-1}, 
one can get the mean values 
\begin{eqnarray}
\label{<A>-qubit}
\langle A\rangle_\rho&=&\widehat{a}\cdot\vec{r}=2p-1\,,
\\
\label{<B>-qubit}
\langle B\rangle_\rho&=&\widehat{b}\cdot\vec{r}=2q-1\,,
\quad\mbox{and}
\\
\label{<C>-qubit}
\langle C\rangle_\rho&=&\widehat{c}\cdot\vec{r}=2s-1\,,
\end{eqnarray}
where
\begin{equation}
\label{pqs}
p=\langle a|\rho|a\rangle\,,\quad
q=\langle b|\rho|b\rangle\,,\quad\mbox{and}\quad
s=\langle c|\rho|c\rangle\,\quad
\end{equation}
are the probabilities [see \eqref{<A>-pro} and \eqref{p-const2}] associated (with $+1$ eigenvalue) to the three projective measurements.
The probabilities $p,q,$ and $s$ are the mean values of three rank-1 projectors
\begin{equation}
\label{proj}
|a\rangle\langle a|=:P\,,\quad
|b\rangle\langle b|=:Q\,,\quad\mbox{and}\quad
|c\rangle\langle c|\,.\quad
\end{equation}

\begin{figure}[]
	\centering
	\includegraphics[width=0.45\textwidth]{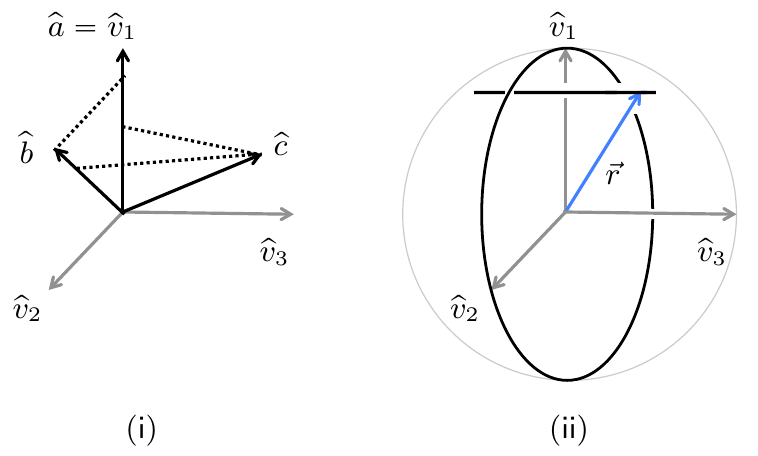}
	\caption{(\textsf{i}) depicts linearly independent unit vectors 
		${\widehat{a},\widehat{b},\widehat{c}}$
		and the orthonormal set 
		${\{\widehat{v}_1,\widehat{v}_2,\widehat{v}_3\}\in\mathbb{R}^3}$.
		There, a dotted line illustrates an orthogonal projection of one vector onto another, which is an integral part of the Gram-Schmidt process.
		(\textsf{ii}) exhibits the Bloch vector $\vec{r}$, a line segment parallel to $\widehat{v}_3$ in the Bloch sphere, and
		the great circle in $v_1v_2$-plane.
	}
	\label{fig:abc-vectors}  
\end{figure}

By applying the Gram-Schmidt orthonormalization process, we can turn the 
linearly independent set ${\{\widehat{a},\widehat{b},\widehat{c}\,\}}$
into an orthonormal set ${\{\widehat{v}_1,\widehat{v}_2,\widehat{v}_3\}}$ of vectors; they are portrayed in Fig.~\ref{fig:abc-vectors}~(\textsf{i}).
The two sets are related through the transformation
\begin{equation}
\label{abc-v1v2v3}
\begin{pmatrix}
\widehat{v}_1\\
\widehat{v}_2\\
\widehat{v}_3
\end{pmatrix}
= 
\underbrace{\begin{pmatrix}
	1 & 0 & 0  \\
	-\tfrac{\widehat{a}.\widehat{b}}
	{\scriptstyle\sqrt{1-(\widehat{a}.\widehat{b})^2}}&
	\tfrac{1}
	{\scriptstyle\sqrt{1-(\widehat{a}.\widehat{b})^2}} & 0   \\
	-\tfrac{e}{g} & -\tfrac{f}{g} & \tfrac{1}{g}  \\ 
	\end{pmatrix}}_{\displaystyle\textbf{M}^{-1}}
\begin{pmatrix}
\widehat{a}\\
\widehat{b}\\
\widehat{c}
\end{pmatrix}\,,
\end{equation}
where
\begin{eqnarray}
\label{e}
e&=&\tfrac{\widehat{a}.\widehat{c}-
	(\widehat{a}.\widehat{b})(\widehat{b}.\widehat{c})}
{1-(\widehat{a}.\widehat{b})^2}\,,
\\
\label{f}
f&=&\tfrac{\widehat{b}.\widehat{c}-
	(\widehat{a}.\widehat{b})(\widehat{a}.\widehat{c})}
{1-(\widehat{a}.\widehat{b})^2}\,,\quad\mbox{and}
\\
\label{g}
g&=&\sqrt{
	\tfrac{1-(\widehat{a}.\widehat{b})^2-(\widehat{a}.\widehat{c})^2-
		(\widehat{b}.\widehat{c})^2+
		2\,(\widehat{a}.\widehat{b})(\widehat{a}.\widehat{c})
		(\widehat{b}.\widehat{c})}
	{1-(\widehat{a}.\widehat{b})^2}}\,.
\end{eqnarray}
We can convert~\eqref{abc-v1v2v3} into
\begin{equation}
\label{ABC-v1v2v3}
\underbrace{\begin{pmatrix}
	\langle A\rangle\\
	\langle B\rangle\\
	\langle C\rangle
	\end{pmatrix}}_{\displaystyle\textsf{E}}
= 
\underbrace{\begin{pmatrix}
	1 & 0 & 0  \\
	{\scriptstyle\widehat{a}.\widehat{b}}&
	{\scriptstyle\sqrt{1-(\widehat{a}.\widehat{b})^2}} & 0   \\
	e+f({\scriptstyle\widehat{a}.\widehat{b}}) & 
	f{\scriptstyle\sqrt{1-(\widehat{a}.\widehat{b})^2}} & 
	g  \\ 
	\end{pmatrix}}_{\displaystyle\textbf{M}}
\underbrace{\begin{pmatrix}
	\widehat{v}_1.\vec{r}\ \\
	\widehat{v}_2.\vec{r}\ \\
	\widehat{v}_3.\vec{r}\
	\end{pmatrix}}_{\displaystyle\texttt{R}}\,,
\end{equation}
which is like Eq.~\eqref{expt-values}.
One can perceive that $\texttt{R}$ is real and it is the representation of Bloch vector $\vec{r}$ in the $v$-coordinate system (made of ${\widehat{v}_1,\widehat{v}_2,\widehat{v}_3}$) [see Fig.~\ref{fig:abc-vectors} (\textsf{ii})].
From top to bottom, the rows in $\textbf{M}$ are the representations of $\widehat{a}$, $\widehat{b}$, and $\widehat{c}$ in the $v$-coordinate system.
Next, one can verify that
\begin{equation}
\label{Gram-matrix}
\textbf{M}\,\textbf{M}^\intercal=
\begin{pmatrix}
1 & {\scriptstyle\widehat{a}.\widehat{b}} & {\scriptstyle\widehat{a}.\widehat{c}}  
\\
{\scriptstyle\widehat{a}.\widehat{b}}&
1 & {\scriptstyle\widehat{b}.\widehat{c}}   \\
{\scriptstyle\widehat{a}.\widehat{c}} & {\scriptstyle\widehat{b}.\widehat{c}} & 1 \\ 
\end{pmatrix}
=:\textbf{G}
\end{equation}
is the Gram matrix. Recall that $\intercal$ symbolizes the transpose.

After associating the Pauli operators with the orthonormal vectors as
\begin{equation}
\label{v-sigma}
\widehat{v}_1\cdot\vec{\sigma}:=X\,,\quad
\widehat{v}_2\cdot\vec{\sigma}:=Y\,,\quad\mbox{and}\quad
\widehat{v}_3\cdot\vec{\sigma}:=Z\,,
\end{equation} 
condition \eqref{|r|^2<=1} emerges as
\begin{equation}
\label{|r|^2<=1-V}
r^2=
(\widehat{v}_1\cdot\vec{r}\,)^2+
(\widehat{v}_2\cdot\vec{r}\,)^2+(\widehat{v}_3\cdot\vec{r}\,)^2
=\texttt{R}^\intercal\texttt{R}\leq 1\,.
\end{equation}
And, with the matrix equation
${\textbf{M}^{-1}\textsf{E}=\texttt{R}}$---gained from \eqref{abc-v1v2v3} or \eqref{ABC-v1v2v3}---we achieve the \emph{quadratic} QC 
\begin{eqnarray}
\label{EGE<=1}
\textsf{E}^\intercal
\underbrace{(\textbf{M}^{-1})^\intercal\,\textbf{M}^{-1}}_{\displaystyle\textbf{G}^{-1}}
\textsf{E}
=\texttt{R}^\intercal\texttt{R}\leq 1\,,
\end{eqnarray}
where
\begin{equation}
\label{G^-1}
\textbf{G}^{-1}= 
\tfrac{1}{\text{det}(\textbf{G})}
\begin{pmatrix}
{\scriptstyle 1\,-\,(\widehat{b}.\widehat{c})^2} & {\scriptstyle(\widehat{a}.\widehat{c})(\widehat{b}.\widehat{c})-
	\widehat{a}.\widehat{b}}&
{\scriptstyle(\widehat{a}.\widehat{b})(\widehat{b}.\widehat{c})-\widehat{a}.\widehat{c}}  
\\
{\scriptstyle(\widehat{a}.\widehat{c})(\widehat{b}.\widehat{c})
	-\widehat{a}.\widehat{b}}&
{\scriptstyle 1\,-\,(\widehat{a}.\widehat{c})^2} & {\scriptstyle(\widehat{a}.\widehat{b})(\widehat{a}.\widehat{c})\,-\,\widehat{b}.\widehat{c}}   
\\
{\scriptstyle(\widehat{a}.\widehat{b})(\widehat{b}.\widehat{c})-\widehat{a}.\widehat{c}} & 
{\scriptstyle(\widehat{a}.\widehat{b})(\widehat{a}.\widehat{c})\,-\,\widehat{b}.\widehat{c}} & 
{\scriptstyle 1\,-\,(\widehat{a}.\widehat{b})^2} 
\end{pmatrix}
\end{equation}
and ${\text{det}(\textbf{G})=({\scriptstyle 1\,-\,(\widehat{a}.\widehat{b})^2})\,g^2}$ [for $g$, see \eqref{g}].
$\textbf{G}^{-1}$ does exist for linearly independent vectors ${\widehat{a},\widehat{b},\widehat{c}}$, otherwise see 
Appendix~\ref{subsec:2settings}.
One can observe that the matrices \textbf{M} and \textbf{G} are independent of $\rho$ and only depend on the three operators (measurement settings).

The quadratic QC in \eqref{EGE<=1} characterizes the
permissible region $\mathcal{E}$ [defined in \eqref{set-of-expt}] of expectation values~\eqref{<A>-qubit}--\eqref{<C>-qubit}.
The linear transformation in~\eqref{ABC-v1v2v3} maps the Bloch sphere identified by the equality in \eqref{|r|^2<=1-V} onto an ellipsoid \cite{Meyer00}.
So, for a qubit, the allowed region $\mathcal{E}$ will always be 
an ellipsoid with its interior \cite{Kaniewski14}.
We want to emphasize that all the material between \eqref{abc-v1v2v3}  and \eqref{G^-1} is given in a general form in \cite{Kaniewski14,Kaniewski,Meyer00}.
It is shown in \cite{Kaniewski14} that there is a one-to-one correspondence between a qubit's state $\rho\in\mathcal{S}$ [defined in \eqref{set-of-states}] and a point in $\mathcal{E}$ as long as \textbf{M} is full rank. That can be witnessed through Eq.~\eqref{ABC-v1v2v3}.

The ellipsoid can be parametrized by putting
\begin{equation}
\label{Rpure}
\texttt{R}^\intercal_{\text{pure}}=(\sin2\theta\cos\phi\,,\
\sin2\theta\sin\phi\,,\
\cos2\theta)
\end{equation}
in \eqref{ABC-v1v2v3}, where ${\theta\in[0,\tfrac{\pi}{2}]}$ and ${\phi\in[0,2\pi)}$.
If we put ${r\texttt{R}^\intercal_{\text{pure}}}$ in \eqref{ABC-v1v2v3}---where ${r\in[0,1]}$ is given in \eqref{|r|^2<=1-V}---then we can also reach its interior points.
The column vector $\texttt{R}_{\text{pure}}$ is associated with
$\texttt{R}^{(\text{pure})}_2$ of
\eqref{R-pure-2}.
For this section,
the subscripts of ${\theta_0}$ and ${\phi_1}$ are dropped.

The real symmetric matrix $\textbf{G}$ can be diagonalized with an orthogonal matrix \textbf{O}, hence 
${\textbf{O}^\intercal \textbf{G}\, \textbf{O}}$ will be a diagonal matrix with entries $\lambda_1$, $\lambda_2$, and $\lambda_3$ at its main diagonal, which are the eigenvalues of \textbf{G}.
The same \textbf{O} also diagonalizes $\textbf{G}^{-1}$, and $\lambda^{-1}_l$ (${l=1,2,3}$) will be its eigenvalues.
With the orthogonal matrix, we can recast condition~\eqref{EGE<=1} 
as
\begin{eqnarray}
\label{ellipsoid-f}
&&
\qquad\quad\frac{{t_1}^2}{\lambda_1}+\frac{{t_2}^2}{\lambda_2}+
\frac{{t_3}^2}{\lambda_3}\leq 1\,,\quad\mbox{where}\\
&&
\label{Ogf}
\textbf{O}^\intercal\,\textsf{E}=
\begin{pmatrix}
t_1\\
t_2\\
t_3
\end{pmatrix}
:=\begin{pmatrix}
\sqrt{\lambda_1}\,\sin\mu\,\cos\nu\\
\sqrt{\lambda_2}\,\sin\mu\,\sin\nu\\
\sqrt{\lambda_3}\,\cos\mu
\end{pmatrix}.
\end{eqnarray}
Through the last equality in \eqref{Ogf}, one can enjoy an alternative parameterization of the ellipsoid, where the parameters ${\mu\in[0,\pi]}$ and ${\nu\in[0,2\pi)}$.
By this technique one can easily find the orientation of the ellipsoid \cite{Meyer00}:
the eigenvectors (that are columns in \textbf{O}) and the eigenvalues $\lambda_i$ of \textbf{G} characterize the semi-principal axes of the ellipsoid.

\subsection{Two measurement settings}\label{subsec:2settings}

In the above investigation, we assume ${\{\widehat{a},\widehat{b},\widehat{c}\}}$ is a set of linearly independent vectors.
Now suppose $\widehat{c}$ is linearly dependent on $\widehat{a}$ and $\widehat{b}$, say $\widehat{c}=\vartheta_a \widehat{a} + 
\vartheta_b \widehat{b} $,
whereas $\widehat{a}$ and $\widehat{b}$ are still linearly independent.
Then, we can discard all the items
related to $\widehat{c}$ in \eqref{ABC-v1v2v3}, and thus achieve an elliptic region $\mathcal{E}$ identified by
\begin{eqnarray}
&&\label{AB-v1v2}
\begin{pmatrix}
2p-1\\
2q-1
\end{pmatrix}=
\underbrace{\begin{pmatrix}
	\langle A\rangle\\
	\langle B\rangle
	\end{pmatrix}}_{\displaystyle\textsf{E}}
= 
\begin{pmatrix}
1 & 0 & 0  \\
{\scriptstyle\widehat{a}.\widehat{b}}&
{\scriptstyle\sqrt{1-(\widehat{a}.\widehat{b})^2}} & 0   \\
\end{pmatrix}
\underbrace{\begin{pmatrix}
	\widehat{v}_1.\vec{r}\ \\
	\widehat{v}_2.\vec{r}\ \\
	\widehat{v}_3.\vec{r}\
	\end{pmatrix}}_{\displaystyle\texttt{R}}\qquad\\
&&
\label{v1v2<1}
\mbox{with}\quad (\widehat{v}_1.\vec{r}\,)^2+(\widehat{v}_2.\vec{r}\,)^2\leq1 \quad \mbox{or}\\
\label{ellipse}
&&\textsf{E}^\intercal\, \textbf{G}^{-1}\textsf{E} \leq 1
\quad \mbox{with}\quad
\textbf{G}= 
\begin{pmatrix}
1 & {\scriptstyle\widehat{a}.\widehat{b}}\\
{\scriptstyle\widehat{a}.\widehat{b}}& 1
\end{pmatrix}.
\end{eqnarray}
We owe \eqref{v1v2<1} and \eqref{ellipse} to \eqref{|r|^2<=1-V}
and \eqref{EGE<=1}, respectively.
The average value $\langle C\rangle=\vartheta_a \langle A\rangle + \vartheta_b \langle B\rangle $ is now just a linear function, 
and the QC, presented by \eqref{AB-v1v2}--\eqref{ellipse}, has no effect of $C$.
To present the QCs, it is sufficient to consider only (linearly) independent 
operators \cite{Adagger}.
So we are ignoring $C$ until Appendix~\ref{subsec:3settings}.

One can notice two things with Eq.~\eqref{AB-v1v2}.
First, a whole line segment---that is in the Bloch sphere and parallel to $\widehat{v}_3$ [displayed in Fig.~\ref{fig:abc-vectors}~(\textsf{ii})]---gets mapped onto a single point in $\mathcal{E}$ under the transformation in \eqref{AB-v1v2}.
Second, extreme points---that constitute the ellipse---of $\mathcal{E}$
come from the pure states that lie on the great circle [illustrated in Fig.~\ref{fig:abc-vectors}~(\textsf{ii})] of the Bloch sphere in the $v_1v_2$-plane.

\begin{figure}
	\centering
	\includegraphics[width=0.45\textwidth]{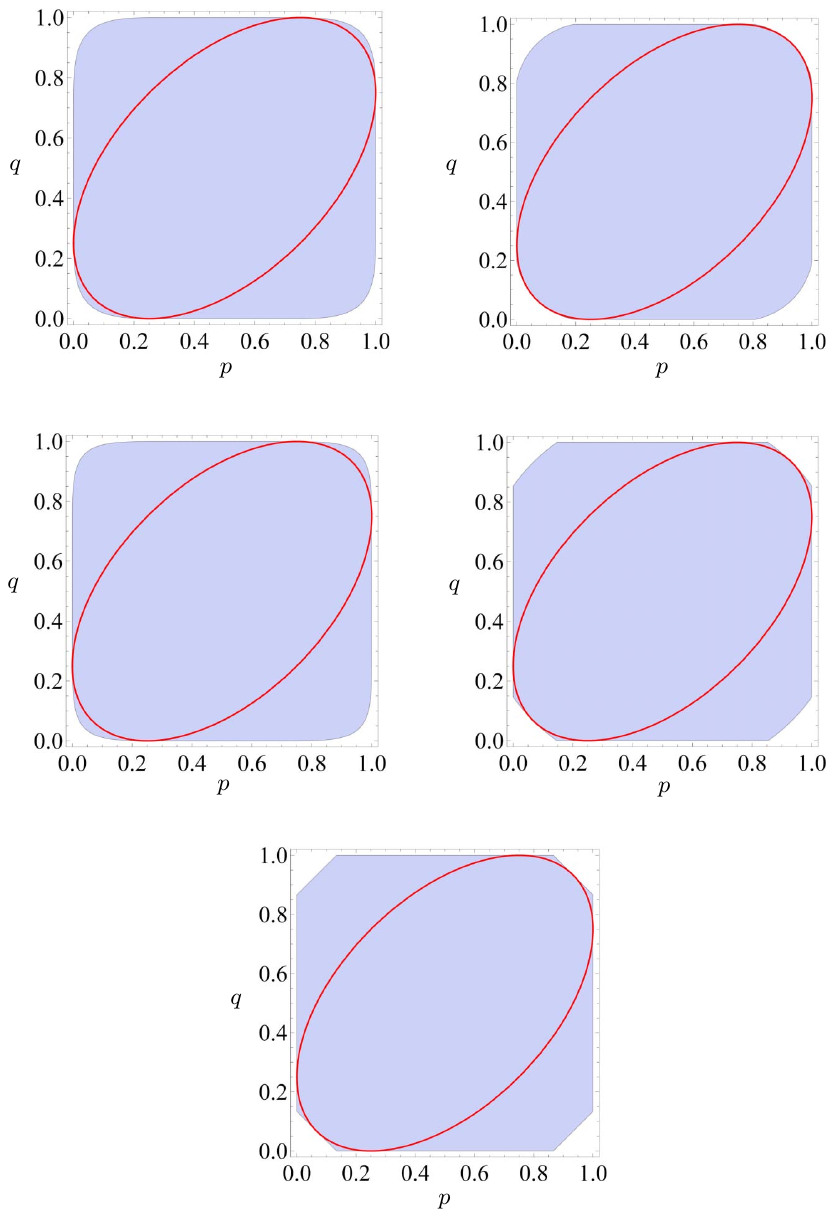}
	\caption{Moving horizontally from top-left to the bottom, the (blue) shaded regions are
		$\mathcal{R}_\Delta$ [defined in~\eqref{region-std-UR}], $\mathcal{R}_H$, $\mathcal{R}_{u_{\sfrac{1}{2}}}$, $\mathcal{R}_{u_2}$, and $\mathcal{R}_{u_\text{max}}$, in that order.
		For each plot, ${\epsilon=\tfrac{3}{4}}$ (that is, ${\widehat{a}\cdot\widehat{b}=\tfrac{1}{2}}$) is taken,
		and the horizontal and vertical axes represent ${p\in[0,1]}$ and ${q\in[0,1]}$, respectively.
		All these $\mathcal{R}$'s contain permissible region~\eqref{ellipse-pq}, which is bounded by the ellipse displayed in each plot.
		The ellipse touches the boundary of a region $\mathcal{R}$ at certain points, some of them correspond to those pure states that saturate the associated CR or UR.
	}
	\label{fig:regions} 
\end{figure}

Equivalently, one can take the projectors $P$ and $Q$ from \eqref{proj} at the places of $A$ and $B$ and then present everything in terms of the probabilities $p$ and $q$ given in \eqref{<A>-qubit}, \eqref{<B>-qubit}, and \eqref{pqs}.
In the case of projectors, hyperrectangle~\eqref{hyperrectangle} becomes the square
${[0,1]^{\times 2}}$, and the allowed region can be described as
\begin{equation}
\label{ellipse-pq}
\mathcal{E}=\{(p,q)\ |\ 0\leq p,q\leq1\ \ \text{obeys}\ \
\eqref{ellipse}\}\,.
\end{equation}
One can check that ${P=\mathring{A}}$ here [for $\mathring{A}$, see \eqref{Adot}], thus ${H(p)=H(\langle A\rangle)}$, which is in fact true for all the uncertainty and certainty measures in~\eqref{H(A)}--\eqref{umax(A)}.

It is shown in \cite{Sehrawat17} that many tight CRs and URs known from \cite{Larsen90,Busch14-b,Garrett90,Sanchez-Ruiz98,Ghirardi03,Bosyk12,Vicente05,Zozor13,Deutsch83,Maassen88,Rastegin12} can be derived by using
ellipse \eqref{AB-v1v2}, and the same ellipse emerges in \cite{Lenard72,Larsen90,Kaniewski14,Abbott16,Sehrawat17} through different methods.
A few such relations are 
\begin{eqnarray}
\label{std-UR}
\sqrt{1-(2\epsilon-1)^2}&\leq&
\Delta P+\Delta Q\,,\\
\label{entropy-UR}
\mbox{if}\ 0.7\leq\epsilon\ \mbox{then}\quad 
2\,h(\tfrac{1+\sqrt{\epsilon}}{2})&\leq&H(p)+H(q)\,,
\\
\label{u-UR}
1+\sqrt{\epsilon}+\sqrt{1-\epsilon}
&\leq&u_{\sfrac{1}{2}}(p)+
u_{\sfrac{1}{2}}(q)\,,\\
\label{u2-CR}
\max\{2-\epsilon,1+\epsilon\}&\geq&u_2(p)+u_2(q)\,,
\ \mbox{and}\quad \\
\label{umax-CR}
\max\big\{1+\sqrt{1-\epsilon}\,,\,1+\sqrt{\epsilon}\,\big\}
&\geq&u_{\textrm{max}}(p)+u_{\textrm{max}}(q)\,,\qquad\quad
\end{eqnarray}
where 
all the above functions are defined according to \eqref{std-dev}--\eqref{umax(A)} for $P$ and $Q$,
\begin{equation}
\label{entropy}
h(p):=-(p\ln p+({1-p})\ln(1-p))\,,\\
\end{equation}
and
${\widehat{a}\cdot\widehat{b}=2\epsilon-1}$.
The standard deviation $\Delta$, the Shannon entropy $H$ \cite{Shannon48}, and $u_{\sfrac{1}{2}}$ are concave functions that provide tight URs \eqref{std-UR}--\eqref{u-UR}, whereas
the convex functions $u_2$ and $u_{\textrm{max}}$ give tight CRs~\eqref{u2-CR} and \eqref{umax-CR} \cite{Sehrawat17}.

Following \eqref{R_H}, one can define a region
\begin{eqnarray}
\label{region-std-UR}
\mathcal{R}_\Delta=\{(p,q)\ |\ 0\leq p,q\leq1\ \ \text{obeys}\ \
\eqref{std-UR}\}\qquad
\end{eqnarray}
that is limited by UR \eqref{std-UR}.
Similarly, one can bound $\mathcal{R}_H$, $\mathcal{R}_{u_{\sfrac{1}{2}}}$, $\mathcal{R}_{u_2}$, and $\mathcal{R}_{u_\text{max}}$ by tight relations \eqref{entropy-UR}--\eqref{umax-CR}.
Taking ${\epsilon=\tfrac{3}{4}}$, we display these regions
in Fig.~\ref{fig:regions} and realize that
${\mathcal{R}_\Delta\subset\mathcal{R}_{u_{\sfrac{1}{2}}}}$ and 
${\mathcal{R}_H\subset \mathcal{R}_{u_2}\subset
	\mathcal{R}_{u_\text{max}}}$, which may or may not hold for other $\epsilon$'s.
Whereas neither
$\mathcal{R}_\Delta$ is a subset of $\mathcal{R}_H$ nor vice versa.

One can also observe that ${(\epsilon,1)\in\mathcal{E}}$ while ${(1-\epsilon,1)\notin\mathcal{E}}$ in Fig.~\ref{fig:regions}.
In these points, $\epsilon$ and ${1-\epsilon}$ are associated with 
the two distinct probability-vectors ${\vec{p}=(\epsilon,1-\epsilon)}$ and ${\vec{p}\,'=(1-\epsilon,\epsilon)}$, respectively.
After the permutation, ${\vec{p}}$ turns into 
$\vec{p}\,'$ that is forbidden.
It is a distinguish feature of a \emph{quantum} probability 
${p_l=\langle a_l|\rho |a_l\rangle}$ [see \eqref{<A>-pro} and \eqref{p-const2}] 
that ${p_l}$
is not only associated with the measurement setting $a$ but also with the label $l$ for an outcome.

\subsection{Three measurement settings}\label{subsec:3settings}

Let us start with Schr\"{o}dinger's UR~\cite{Schrodinger32}
\begin{eqnarray}
\label{Schrodinger-UR}
&&
{\scriptstyle 0\,\leq\, \left(\langle A^2\rangle-\langle A\rangle^2\right)\left(
	\langle B^2\rangle-\langle B\rangle^2\right)
	-|\langle \widetilde{C}\rangle|^2
	-|\langle \widetilde{D}\rangle-\langle A\rangle\langle B\rangle|^2}\,,
\nonumber\\
&&
\mbox{where}\quad
{\scriptstyle\widetilde{C}\,:=\frac{AB-BA}{2\,\text{i}}\quad\mbox{and}\quad
	\widetilde{D}\,:=\frac{AB+BA}{2}}\quad
\end{eqnarray}
are related to the commutator and the anticommutator, respectively, of $A$ and $B$.
For qubit's operators~\eqref{A-qubit} and \eqref{B-qubit}, one can realize through~\eqref{AB} that
\begin{equation}
\label{c=a cros b}
\widetilde{C}=(\widehat{a}\times\widehat{b})\cdot\vec{\sigma}=
|\widehat{a}\times\widehat{b}|
\underbrace{\widehat{c}\cdot\vec{\sigma}}_{C}\,,\quad
\widehat{c}=\frac{\widehat{a}\times\widehat{b}}{|\widehat{a}\times\widehat{b}|}\,,
\end{equation} 
${\widetilde{D}=(\widehat{a}\cdot\widehat{b})\,I}$, and ${A^2=I=B^2}$.
Considering these and 
${\scriptstyle|\widehat{a}\times\widehat{b}|=\sqrt{1-(\widehat{a}\cdot\widehat{b})^2}}$,
we can rewrite Schr\"{o}dinger's UR for a qubit as
\begin{equation}
\label{Schrodinger-UR-qubit}
0\leq (1-\langle A\rangle^2)(
1-\langle B\rangle^2)
-({\scriptstyle 1-(\widehat{a}\cdot\widehat{b})^2})\langle C\rangle^2
-({\scriptstyle\widehat{a}\cdot\widehat{b}}\,-\langle
A\rangle\langle B\rangle)^2.
\end{equation}
To test \eqref{Schrodinger-UR-qubit} in experimental scenario~\eqref{expt-situ}, one requires three measurement settings
${\widehat{a},\widehat{b},\widehat{c}}$.
One can choose $\widehat{a}$ and $\widehat{b}$, and then $\widehat{c}$
is fixed by the cross product in~\eqref{c=a cros b}.
If one takes $\widehat{a}$ and $\widehat{b}$ collinear, then
\eqref{Schrodinger-UR-qubit} turns into the trivial statement ${0=0}$. 
So we are taking $\widehat{a}$ and $\widehat{b}$ linearly independent.

\begin{figure}[ ]
	\centering
	\includegraphics[width=0.30\textwidth]{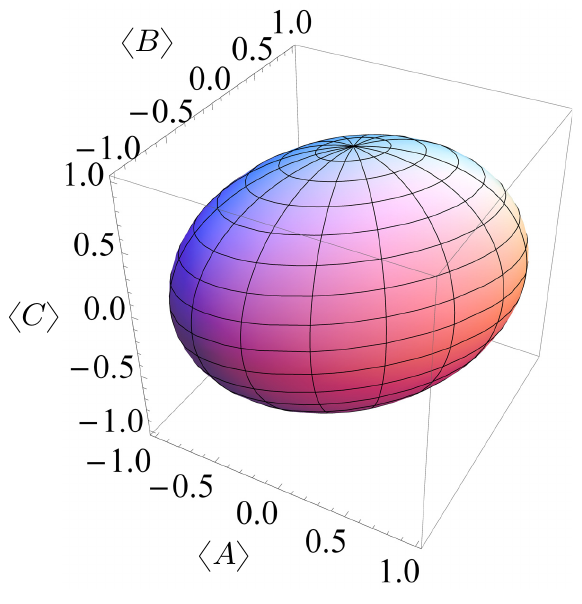} 
	\caption{The region of ${\langle A\rangle,\langle B\rangle,\langle C\rangle\in[-1,1]}$, which is restricted by Schr\"{o}dinger's UR~\eqref{Schrodinger-UR-qubit}.
		It is, ellipsoidal in shape, presented by picking ${\widehat{a}\cdot\widehat{b}=\tfrac{1}{2}}$, and $C$ is determined by \eqref{c=a cros b}.
		The ellipsoid turns into the Bloch sphere for orthogonal 
		$\widehat{a}$ and $\widehat{b}$.
	}
	\label{fig:ellip-SUR}  
\end{figure}

One can check that Schr\"{o}dinger's UR~\eqref{Schrodinger-UR-qubit}
and QC \eqref{EGE<=1} with the Gram matrix
\begin{equation}
\label{SR-Gram-matrix}
\textbf{G}_{\text{Sch}}= 
\begin{pmatrix}
1 & {\scriptstyle\widehat{a}\cdot\widehat{b}} & 0 \\[1mm]
{\scriptstyle\widehat{a}\cdot\widehat{b}} & 1 & 0 \\[1mm]
0 &0 & 1 \\ 
\end{pmatrix}
\end{equation}
are the same thing, and the UR is saturated by every pure state for a qubit.
Without the last term in \eqref{Schrodinger-UR-qubit}, Schr\"{o}dinger's UR becomes Robertson's UR \cite{Robertson29}, which will form a bigger region
than the allowed region here characterized by \eqref{Schrodinger-UR-qubit}.

Taking ${\widehat{a}\cdot\widehat{b}=\tfrac{1}{2}}$,
the ellipsoid is displayed in Fig.~\ref{fig:ellip-SUR}.
Orthogonal projection of the ellipsoid onto the 
${\langle A\rangle\langle B\rangle}$--plane produces the same elliptic region that is identified by~\eqref{ellipse} and shown in Fig.~\ref{fig:regions}.
The parametric forms [obtained via~\eqref{Ogf} and \eqref{ABC-v1v2v3} with \eqref{Rpure}] of the ellipsoid are
\begin{eqnarray}
\label{S-UR-ellipsoid-para}
\underbrace{\begin{pmatrix}
	\langle A\rangle\\
	\langle B\rangle\\
	\langle C\rangle
	\end{pmatrix}}_{\displaystyle\textsf{E}}
&=& 
\underbrace{\begin{pmatrix}
	\tfrac{1}{\sqrt{2}} & \hphantom{-}\tfrac{1}{\sqrt{2}} & 0 \\[1mm]
	\tfrac{1}{\sqrt{2}} & -\tfrac{1}{\sqrt{2}} & 0 \\[1mm]
	0 & \hphantom{-}0 & 1 \\ 
	\end{pmatrix}}_{\displaystyle\textbf{O}}
\begin{pmatrix}
{\scriptstyle\sqrt{1+\widehat{a}\cdot\widehat{b}}}\,\sin\mu\,\cos\nu\\
{\scriptstyle\sqrt{1-\widehat{a}\cdot\widehat{b}}}\,\sin\mu\,\sin\nu\\
\quad \cos\mu
\end{pmatrix}\qquad\\
&=&
\underbrace{\begin{pmatrix}
	1 & 0 & 0 \\[1mm]
	{\scriptstyle\widehat{a}\cdot\widehat{b}} & {\scriptstyle\sqrt{1-(\widehat{a}\cdot\widehat{b})^2}} & 0 \\[1mm]
	0 & 0  & 1 \\ 
	\end{pmatrix}}_{\displaystyle\textbf{M}}
\underbrace{\begin{pmatrix}
	\sin2\theta\,\cos\phi\\
	\sin2\theta\,\sin\phi\\
	\cos2\theta
	\end{pmatrix}}_{\displaystyle\texttt{R}_\text{pure}}\ .
\end{eqnarray}
One can easily recognize the semi-principal axes in Fig.~\ref{fig:ellip-SUR} with \eqref{S-UR-ellipsoid-para}.

For the next example, we consider three linearly-independent unit vectors
${\widehat{\mathsf{a}}_1,\widehat{\mathsf{a}}_2,\widehat{\mathsf{a}}_3}$ such that their 
Gram matrix is
\begin{equation}
\label{SIC-POVM-Gram-matrix}
\textbf{G}_{\textsc{sic}}= 
\begin{pmatrix}
\hphantom{-}1 & -\tfrac{1}{3} & -\tfrac{1}{3} \\[1mm]
-\tfrac{1}{3} & \hphantom{-}1 & -\tfrac{1}{3} \\[1mm]
-\tfrac{1}{3} & -\tfrac{1}{3} & \hphantom{-}1 \\ 
\end{pmatrix}\,.
\end{equation}
It implies that there is an equal angle, $\arccos(-\tfrac{1}{3})$, between every pair of the vectors. There exists one more such unit vector ${\widehat{\mathsf{a}}_4=-\textstyle\sum\nolimits_{i=1}^{3}
	\widehat{\mathsf{a}}_i}$.
The set of four vectors 
${\{\widehat{\mathsf{a}}_i\}_{i=1}^{4}}$
yields a SIC-POVM for a qubit \cite{Rehacek04,Appleby09},
whose elements are the positive semi-definite operators
\begin{equation}
\label{Pi}
\varPi_i=\tfrac{1}{4}(I+\widehat{\mathsf{a}}_i\cdot\vec{\sigma})\,,
\quad\mbox{and}\quad
\sum_{i=1}^{4}\varPi_i=I
\end{equation}
is because $\textstyle\sum\nolimits_{i=1}^{4}\widehat{\mathsf{a}}_i$ is a null vector.

\begin{figure}
	\centering
	\includegraphics[width=0.45\textwidth]{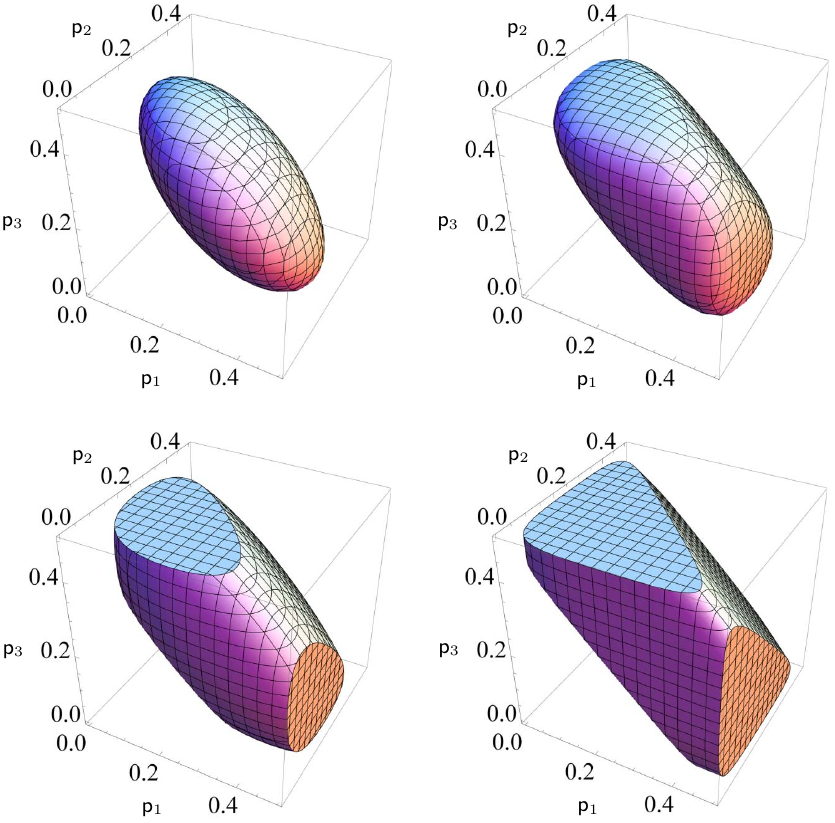}
	\caption{From top-left to bottom-right, moving horizontally, the first one is the allowed (ellipsoidal) region $\mathcal{E}$ restrained by QCs \eqref{Pi=1} and \eqref{Pi<=1/3}. The second, third, and fourth regions are $\mathcal{R}^{\textsc{sic}}_\Delta$, $\mathcal{R}^{\textsc{sic}}_h$, and $\mathcal{R}^{\textsc{sic}}_{\mathsf{u}_{\sfrac{1}{2}}}$, respectively, which are described in the main text. 
	$\mathcal{E}$ lies within every $\mathcal{R}^{\textsc{sic}}$
	and shares a set of boundary points, which come from the minimum uncertainty states that saturate the associated UR.
	One can also observe that 
	${\mathcal{R}^{\textsc{sic}}_\Delta\subset\mathcal{R}^{\textsc{sic}}_{\mathsf{u}_{\sfrac{1}{2}}}}$ and ${\mathcal{R}^{\textsc{sic}}_h\subset\mathcal{R}^{\textsc{sic}}_{\mathsf{u}_{\sfrac{1}{2}}}}$.
	}
	\label{fig:regions-SIC-POVM} 
\end{figure}

Since eigenvalues of every $\varPi_i$ are $0$ and $\tfrac{1}{2}$,
its mean value
${\langle \varPi_i\rangle:=\mathsf{p}_i\in[0,\tfrac{1}{2}]}$ according to \eqref{<A>in}. 
Moreover, for ${(\varPi_1,\varPi_2,\varPi_3)}$, hyperrectangle~\eqref{hyperrectangle} is the cube ${[0,\tfrac{1}{2}]^{\times 3}}$ in which regions are exhibited in Fig.~\ref{fig:regions-SIC-POVM}.
The linear QC 
\begin{equation}
\label{Pi=1}
\sum_{i=1}^{4}\,
\langle\widehat{\mathsf{a}}_i\cdot\vec{\sigma}\rangle=0
\quad\Leftrightarrow\quad
\sum_{i=1}^{4}\mathsf{p}_i=1
\end{equation}
is due to normalization~\eqref{norm-rho} of a state, that is, ${\langle I\rangle=1}$, where the identity operator is given in \eqref{Pi}.
To estimate ${\langle\widehat{\mathsf{a}}_i\cdot\vec{\sigma}\rangle=4\,\mathsf{p}_i-1}$, ${i=1,\cdots,4}$, 
one can either choose three projective measurements along $\widehat{\mathsf{a}}_1, \widehat{\mathsf{a}}_2$, and $\widehat{\mathsf{a}}_3$ or the single POVM ${\{\varPi_i\}_{i=1}^{4}}$ that can be realized by the scheme in \cite{Rehacek04}.
In either case, the permitted region $\mathcal{E}$ 
is identified by \eqref{Pi=1}
and the quadratic QC
\begin{equation}
\label{Pi<=1/3}
\sum_{i=1}^{4}\,
\langle\widehat{\mathsf{a}}_i\cdot\vec{\sigma}\rangle^2\leq\tfrac{4}{3}
\quad\Leftrightarrow\quad
\sum_{i=1}^{4}\mathsf{p}_i^2
\leq\tfrac{1}{3}\,.
\end{equation}
The right-hand-side inequality is given in \cite{Rehacek04}, and it can also be derived from~\eqref{EGE<=1} by using Gram matrix \eqref{SIC-POVM-Gram-matrix}.
Employing~\eqref{Ogf} and \eqref{ABC-v1v2v3} with \eqref{Rpure}, we can have parametric forms
\begin{eqnarray}
\label{SIC-POVM-ellipsoid-para-O}
\underbrace{\begin{pmatrix}
	4\,\mathsf{p}_1-1\\
	4\,\mathsf{p}_2-1\\
	4\,\mathsf{p}_3-1
	\end{pmatrix}}_{\displaystyle\textsf{E}}
&=& 
\underbrace{\begin{pmatrix}
	\tfrac{1}{\sqrt{3}} & \hphantom{-}\tfrac{1}{\sqrt{2}} & -\tfrac{1}{\sqrt{6}} \\[1mm]
	\tfrac{1}{\sqrt{3}} & -\tfrac{1}{\sqrt{2}} & -\tfrac{1}{\sqrt{6}} \\[1mm]
	\tfrac{1}{\sqrt{3}} & \hphantom{-}0 & \hphantom{-}\tfrac{2}{\sqrt{6}} \\ 
	\end{pmatrix}}_{\displaystyle\textbf{O}}
\begin{pmatrix}
\tfrac{1}{\sqrt{3}}\,\sin\mu\,\cos\nu\\
\tfrac{2}{\sqrt{3}}\,\sin\mu\,\sin\nu\\
\tfrac{2}{\sqrt{3}}\,\cos\mu
\end{pmatrix}\qquad\ \ \\
\label{SIC-POVM-ellipsoid-para-M}
&=&
\label{M^-1 SIC}
\underbrace{\begin{pmatrix}
	\hphantom{-}1 & \hphantom{-}0 & 0 \\[1mm]
	-\tfrac{1}{3} & \hphantom{-}\tfrac{2\sqrt{2}}{3} & 0 \\[1mm]
	-\tfrac{1}{3} & -\tfrac{\sqrt{2}}{3}  & \tfrac{\sqrt{2}}{\sqrt{3}} \\ 
	\end{pmatrix}}_{\displaystyle\textbf{M}}
\underbrace{\begin{pmatrix}
	\sin2\theta\,\cos\phi\\
	\sin2\theta\,\sin\phi\\
	\cos2\theta
	\end{pmatrix}}_{\displaystyle\texttt{R}_\text{pure}}.
\end{eqnarray}
of the ellipsoid that is the boundary of $\mathcal{E}$.
Taking \eqref{SIC-POVM-ellipsoid-para-O}, one can check orientation of the ellipsoid exhibited in Fig.~\ref{fig:regions-SIC-POVM} (top-left).

\begin{figure}[ ]
	\centering
	\includegraphics[width=0.35\textwidth]{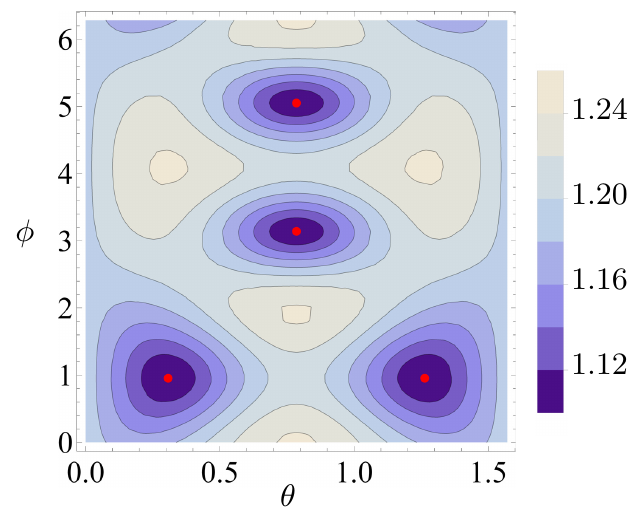}
	\caption{Contour plot of the entropy $-\textstyle\sum\nolimits_{i=1}^{4}\mathsf{p}_i\ln\mathsf{p}_i$, which reaches its global minimum ${\ln 3}$ at the four (red) points that are associated with ${\{-\widehat{\mathsf{a}}_i\}_{i=1}^{4}}$. 
	}
	\label{fig:entropy-plot-SIC} 
\end{figure}

\begin{table}[H]
	\centering
	\caption{The
		values of $\theta$ and $\phi$ for  ${\{-\widehat{\mathsf{a}}_i\}_{i=1}^{4}}$ are drawn according to
		\eqref{Rpure}, which parameterizes every unit vector in $\mathbb{R}^3$.
		By replacing $2\theta$ and $\phi$ with ${\pi-2\theta}$ and ${\pi+\phi}$, respectively, one can get the values for the antipodal vectors ${\{\widehat{\mathsf{a}}_i\}_{i=1}^{4}}$.  
	}
	\label{tab:a-theta-phi}
	\begin{tabular}{c@{\hspace{2mm}} | @{\hspace{2mm}}c@{\hspace{3mm}} c }
		\hline\hline\rule{0pt}{2ex}  
		& $2\theta$ & $\phi$  \\
		\hline\rule{0pt}{3ex} 	
		$-\,\widehat{\mathsf{a}}_1$ & $\tfrac{\pi}{2}$ & $\pi$
		\\[1mm]
		$-\,\widehat{\mathsf{a}}_2$ & 
		$\tfrac{\pi}{2}$ & $\pi+\arccos(-\tfrac{1}{3})$
		\\[1mm]
		$-\,\widehat{\mathsf{a}}_3$ & 
		$\pi-\arccos(\tfrac{\sqrt{2}}{\sqrt{3}})$ & $\hphantom{\pi+}\arccos(\tfrac{1}{\sqrt{3}})$ 
		\\[1mm]
		$-\,\widehat{\mathsf{a}}_4$ & $\hphantom{\pi-}\arccos(\tfrac{\sqrt{2}}{\sqrt{3}})$ & $\hphantom{\pi+}\arccos(\tfrac{1}{\sqrt{3}})$   \\[1.2mm]			
		\hline\hline
	\end{tabular}
	\label{tab:theta-phi-for(-a)}
\end{table}

To measure a combined uncertainty,
if one picks a suitable concave function of ${\{\mathsf{p}_i\}_{i=1}^{4}}$,
for example, the standard Shannon entropy $-\textstyle\sum\nolimits_{i=1}^{4}\mathsf{p}_i\ln\mathsf{p}_i$,
then its absolute minimum will occur on the ellipsoid 
parametrized by $\theta$ and $\phi$ in \eqref{SIC-POVM-ellipsoid-para-M}.
By plotting the entropy as a function
of ${\theta\in[0,\tfrac{\pi}{2}]}$ and ${\phi\in[0,2\pi)}$ in Fig.~\ref{fig:entropy-plot-SIC},
we observe that the entropy reaches its absolute minimum
${\ln3}$ when the Bloch vector ${\vec{r}\in\{-\widehat{\mathsf{a}}_i\}_{i=1}^{4}}$;
the corresponding $\theta$'s and $\phi$'s are registered in 
Table~\ref{tab:a-theta-phi}.
In this way, we establish three tight URs 
\begin{eqnarray}
\label{std-SIC}
2\sqrt{2}&\,\leq\,&
\sum_{i=1}^{4}
\sqrt{1-\langle\widehat{\mathsf{a}}_i\cdot\vec{\sigma}\rangle^2}
=\sum_{i=1}^{4}\sqrt{1-(4\mathsf{p}_i-1)^2}\,,\qquad\ \ \\
\label{entropy-SIC}
\ln3&\,\leq\,&
-\sum_{i=1}^{4}\mathsf{p}_i\ln\mathsf{p}_i =
	h(\mathsf{p}_1,\cdots,\mathsf{p}_4)\,,
\quad\mbox{and}\\
\label{u-SIC}
\sqrt{3}&\,\leq\,&
\sum_{i=1}^{4}\sqrt{\mathsf{p}_i} =:
	\mathsf{u}_{\sfrac{1}{2}}(\mathsf{p}_1,\cdots,\mathsf{p}_4)\,.
\end{eqnarray}
The right-hand-side is the sum of standard deviations in UR~\eqref{std-SIC}, which is saturated by eight pure states,
whose Block vectors ${\vec{r}\in\{\pm\widehat{\mathsf{a}}_i\}_{i=1}^{4}}$.
Whereas, both URs~\eqref{entropy-SIC} and \eqref{u-SIC} are saturated by four pure states that are related to ${\{-\widehat{\mathsf{a}}_i\}_{i=1}^{4}}$.
Note that the uncertainty measures $h$ and $\mathsf{u}_{\sfrac{1}{2}}$ in \eqref{entropy-SIC} and \eqref{u-SIC} are different from \eqref{H(A)} and \eqref{u(A)}.
Since
${\textstyle\sum\nolimits_{i=1}^{4}\mathsf{p}_i^2}$ is a convex function
\cite{Sehrawat17}, the right-hand-side inequality in \eqref{Pi<=1/3} can be seen as a tight CR.
While the left-hand-side inequality delivers a tight UR for the sum of squared standard deviations ${1-\langle\widehat{\mathsf{a}}_i\cdot\vec{\sigma}\rangle^2}$, and the sum is bounded by $\tfrac{8}{3}$ from below.
Both these relations are saturated by every pure state.

As before, we can restrict a set of ${(\mathsf{p}_1,\cdots,\mathsf{p}_3)}$ by
one of the above URs, for instance,
\begin{eqnarray}
\label{R-std-SIC}
&&\mathcal{R}^{\textsc{sic}}_\Delta:=\{(\mathsf{p}_1,\cdots,\mathsf{p}_3)\ |\ 0\leq \mathsf{p}_1,\cdots,\mathsf{p}_4\leq \tfrac{1}{2}\ \nonumber\\
&& 
\qquad\qquad\qquad\qquad\qquad \text{obey}\
\eqref{Pi=1}\ \text{and}\ \eqref{std-SIC}\}\,.\qquad
\end{eqnarray}
Replacing UR~\eqref{std-SIC} in \eqref{R-std-SIC} by \eqref{entropy-SIC}
and \eqref{u-SIC}, we define the regions $\mathcal{R}^{\textsc{sic}}_h$ and
$\mathcal{R}^{\textsc{sic}}_{\mathsf{u}_{\sfrac{1}{2}}}$, respectively.
One can see cross sections in $\mathcal{R}^{\textsc{sic}}_h$ and
$\mathcal{R}^{\textsc{sic}}_{\mathsf{u}_{\sfrac{1}{2}}}$ caused by ${\mathsf{p}_i=\tfrac{1}{2}}$ ${(i=1,\cdots,4)}$, which
shows a significance of \eqref{<A>in}.





\begin{thebibliography}{99}
\bibitem{+veMat}
All principal minors of a positive semi-definite matrix are nonnegative numbers (see Chap.~7 in the book cited in \cite{Meyer00}). So, one can use these necessary and sufficient conditions---for a valid density matrix made of $\texttt{r}_{jk}$ [given in \eqref{rho-in-stnd-basis}]---as a substitute for the conditions in \eqref{0<=S_n}.
In fact, $S_n$ of \eqref{S_n} is the sum of all ${n\times n}$ principal minors of the density matrix.
\bibitem{unitary-eq}
Picking a different orthonormal basis, than \eqref{HS-basis}, one can get different $\texttt{R}'$ and $\texttt{A}'$.
Since a pair of orthonormal bases is related through a unitary transformation, which does not change the inner product
${\lgroup\rho,A\rgroup_\textsc{hs}=\langle A\rangle_{\rho}}$, 
we reach the same expectation value.
\bibitem{SIC-POVM}
However, one can realize all the mean values 
in $\overline{\texttt{R}}$ in \eqref{expt-values} by employing a single informationally complete POVM \cite{Appleby09}.
To experimentally implement such a POVM, one needs to attach the system of interest to an additional system and then has to perform a projective measurement on the combined system (for example, see \cite{Rehacek04}).
Elements of a POVM---are positive operators that add up to the identity operator---do not commute in general and yet realized with a single measurement setting, of course, on a larger system.
\bibitem{con-to-con} 
One can easily show that a linear transformation carries a convex set onto a convex set.
\bibitem{Adagger} 
There is only one, not two, independent operator in the set ${\{A, A^\dagger\}}$, although their relation is not linear because the adjoint is an antilinear map.
Nonetheless, their average values are related ${\langle A^\dagger\rangle= \overline{\langle A\rangle}}$ [see \eqref{r}]: if we know one, then we know the other.
\bibitem{A}
Instead of $A$ defined in \eqref{A-qubit}, one can take more general Hermitian operator $\widetilde{A}:=a_0I+\vec{a}.\vec{\sigma}$, 
where $\widehat{a}$ is the unit vector of $\vec{a}$. 
Nevertheless, the expectation value of $\widetilde{A}$ can be obtained from 
${\langle A\rangle}$: ${\langle \widetilde{A}\rangle=a_0+\tfrac{1}{|\vec{a}|}\langle A\rangle}$.
So, we do not lose generality if we take $A$ rather than $\widetilde{A}$.
\end{thebibliography}
\end{document}